%% file: main.tex
\documentclass[final,3p,times,12pt]{elsarticle}

\usepackage{amssymb}
\usepackage{tikz}

\usepackage{lineno}

\journal{Nuclear Instruments and Methods A}

\input{testbeam-symbols-def}
\usepackage[capitalise, noabbrev]{cleveref}
\usepackage{siunitx}
\usepackage{etoolbox}
\usepackage{appendix}

\usepackage{todonotes}
\newcommand{\mevneq}{\text{1\xspace MeV\xspace n$_\text{eq}$/cm$^{2}$}}

\input{lhcb-symbols-def}
\usepackage[section]{placeins}
\usepackage[capitalise, noabbrev]{cleveref}
\usepackage{tabularx} 
\begin{document}

\begin{frontmatter}

\vspace*{-3.5cm}
\centerline{\large EUROPEAN ORGANIZATION FOR NUCLEAR RESEARCH (CERN)}
\vspace*{.5cm}
\noindent

\begin{tabular*}{\textwidth}{l @{\extracolsep{\fill}} r}
& LHCb-PUB-2021-010 \\  
 & \today \\ 
 & \\
\end{tabular*}
\vspace*{0.1cm}


\title{Microchannel cooling for the LHCb VELO Upgrade I}


\author[CERN,UOM,*]{Oscar Augusto De Aguiar Francisco}
\author[CERN,KRAKOW,*]{Wiktor~Byczynski}
\author[NIKHEF]{Kazu~Akiba}
\author[CERN,UOM]{Claudia~Bertella}
\author[UOM]{Alexander~Bitadze}
\author[OXFORD]{Matthew~Brock}
\author[CERN]{Bartosz~Bulat}
\author[CERN]{Guillaume~Button}
\author[CERN]{Jan~Buytaert}
\author[UOM]{Stefano~De~Capua}
\author[CERN]{Riccardo~Callegari}
\author[LETI]{Christine~Castellana}
\author[CERN]{Andrea~Catinaccio}
\author[LETI]{Catherine~Charrier}
\author[CERN]{Collette~Charvet}
\author[CERN]{Victor~Coco}
\author[CERN]{Paula~Collins}
\author[CERN]{Jordan~Degrange}
\author[CERN]{Raphael~Dumps}
\author[CERN]{Diego~Alvarez~Feito}
\author[UOM]{Julian~Freestone}
\author[CERN,AGH]{Mariusz~Jedrychowski}  
\author[LIVERPOOL]{Vinicius~Franco~Lima}
\author[SANTIAGO]{Abraham~Gallas}
\author[NIKHEF]{Wouter~Hulsbergen}
\author[NIKHEF]{Daniel~Hynds}
\author[CERN]{Gonzalo~Arnau~Izquierdo} 
\author[OXFORD]{Pawel~Jalocha}
\author[NIKHEF]{Eddy~Jans}
\author[OXFORD]{Malcolm~John}
\author[CERN,OXFORD]{Nathan~Jurik}
\author[SINP]{Alexander~Leflat}
\author[SANTIAGO]{Edgar~Lemos~Cid}
\author[CERN]{Rolf~Lindner}
\author[CERN,EPFL]{Alessandro~Mapelli}
\author[CERN]{Jerome~Noel}
\author[BNL]{Andrei~Nomerotski}
\author[CERN]{Rui~de~Oliveira}
\author[NIKHEF]{Martijn~van~Overbeek}
\author[CERN,UOM]{Chris~Parkes}
\author[CERN]{Paolo~Petagna}
\author[CERN]{Alexandre~Porret}
\author[LETI]{Denis~Renaud}
\author[NIKHEF]{Erno~Roeland}
\author[CERN]{Giulia~Romagnoli}
\author[LETI]{Eric~Rouchouze}
\author[NIKHEF]{Krista~de~Roo}
\author[CERN,NIKHEF]{Freek~Sanders}
\author[CERN]{Thomas~Schneider}
\author[CERN]{Heinrich~Schindler}
\author[CERN]{Burkhard~Schmidt}
\author[CERN]{Andreas~Schopper}
\author[OXFORD]{Luke~Scantlebury-Smead}
\author[CERN]{Miranda~Van~Stenis}
\author[CERN,UOM]{Peter~Svihra}
\author[CERN]{Benoit~Teissandier}
\author[LETI]{Jean-Francois~Teissier}
\author[CERN]{Xavier~Thery}
\author[CERN]{Eric~Thomas}
\author[CERN]{Bart~Verlaat}

\affiliation[CERN]{organization={CERN},
            city={Geneva},
            country={Switzerland}}


\affiliation[UOM]{organization={The University of Manchester},
            city={Manchester},
            country={United Kingdom}}
            
\affiliation[Nikhef]{organization={NIKHEF},
            city={Amsterdam},
            country={Netherlands}}

\affiliation[OXFORD]{organization={University of Oxford},
            city={Oxford},
            country={United Kingdom}}
            
\affiliation[LIVERPOOL]{organization={Liverpool University},
            city={Liverpool},
            country={United Kingdom}}

\affiliation[SANTIAGO]{organization={Universidad de Santiago de Compostela},
            city={Moscow},
            country={Russia}}

\affiliation[SINP]{organization={Institute of Nuclear Physics, Moscow State University},
            city={Santiago de Compostela},
            country={Spain}}
\affiliation[KRAKOW]{organization={Cracow University of Technology},
            city={Krakow},
            country={Poland}}

\affiliation[LETI]{organization={CEA, LETI},
            city={Grenoble},
            country={France}}

\affiliation[EPFL]{organization={EPFL},
            city={Lausanne},
            country={Switzerland}}

\affiliation[BNL]{organization={Brookhaven National Laboratory},
            city={Upton, New York},
            country={United States of America}}

\affiliation[AGH]{organization={AGH University of Science and Technology},
            city={Krakow},
            country={Poland}}
\affiliation[*]{Corresponding authors; oscar.augusto@cern.ch, wiktor.byczynski@cern.ch}
\begin{abstract}
The LHCb VELO Upgrade I, currently being installed for the 2022 start of LHC Run 3, uses silicon microchannel coolers with internally circulating bi-phase 
\cotwo for thermal control of hybrid pixel modules operating in vacuum.  This is the largest scale application of this technology to date.  Production of the microchannel coolers was completed in July 2019 and the assembly into cooling structures was completed in September 2021.  This paper describes the R\&D path supporting the microchannel production and assembly and the motivation for the design choices.  The microchannel coolers have excellent thermal peformance, low and uniform mass, no thermal expansion mismatch with the ASICs and are radiation hard.  The fluidic and thermal performance is presented.
\end{abstract}




\end{frontmatter}


\input{01_introduction}

\input{02_prototyping}


\input{03_LHCb_implementation}

\input{04_performance_simulation}

\input{05_microchannel_assembly}


\input{06_microchannel_qa}


\input{07_cooling_performance}


\input{08_summary}

\appendix


 \bibliographystyle{elsarticle-num} 
 \bibliography{cas-refs}
\biboptions{sort&compress}





\end{document}

%% file: lhcb-symbols-def.tex
\usepackage{xspace} 
\usepackage{upgreek}

\def\MagUp {\mbox{\em Mag\kern -0.05em Up}\xspace}

{

 \def\PDelta      {\ensuremath{\Delta}\xspace}                 
 \def\PXi      {\ensuremath{\Xi}\xspace}                 
 \def\PLambda      {\ensuremath{\Lambda}\xspace}                 
 \def\PSigma      {\ensuremath{\Sigma}\xspace}                 
 \def\POmega      {\ensuremath{\Omega}\xspace}                 
 \def\PUpsilon      {\ensuremath{\Upsilon}\xspace}                 
 
 \def\PB      {\ensuremath{\mathrm{B}}\xspace}                 
                  
 \def\PD      {\ensuremath{\mathrm{D}}\xspace}

 \def\PK      {\ensuremath{\mathrm{K}}\xspace}

 \def\Pi      {\ensuremath{\mathrm{i}}\xspace}

}
{

 \mathchardef\PDelta="7101
 \mathchardef\PXi="7104
 \mathchardef\PLambda="7103
 \mathchardef\PSigma="7106
 \mathchardef\POmega="710A
 \mathchardef\PUpsilon="7107
                  
 \def\PB      {\ensuremath{B}\xspace}                 
                  
 \def\PD      {\ensuremath{D}\xspace}

 \def\PK      {\ensuremath{K}\xspace}

 \def\Pi      {\ensuremath{i}\xspace}

}

\makeatletter
\ifcase \@ptsize \relax
  \newcommand{\miniscule}{\@setfontsize\miniscule{4}{5}}
\or
  \newcommand{\miniscule}{\@setfontsize\miniscule{5}{6}}
\or
  \newcommand{\miniscule}{\@setfontsize\miniscule{5}{6}}
\fi
\makeatother

\DeclareRobustCommand{\optbar}[1]{\shortstack{{\miniscule (\rule[.5ex]{1.25em}{.18mm})}
  \\ [-.7ex] $#1$}}

  \def\Kbar    {{\kern 0.2em\overline{\kern -0.2em \PK}{}}\xspace}

\def\KorKbar    {\kern 0.18em\optbar{\kern -0.18em K}{}\xspace}

  \def\Dbar    {{\kern 0.2em\overline{\kern -0.2em \PD}{}}\xspace}

\def\DorDbar    {\kern 0.18em\optbar{\kern -0.18em D}{}\xspace}

\def\Bbar    {{\ensuremath{\kern 0.18em\overline{\kern -0.18em \PB}{}}}\xspace}

\def\BorBbar    {\kern 0.18em\optbar{\kern -0.18em B}{}\xspace}

  \def\Y#1S{\ensuremath{\PUpsilon{(#1S)}}\xspace}

\def\Lbar        {{\ensuremath{\kern 0.1em\overline{\kern -0.1em\PLambda}}}\xspace}
\def\LorLbar    {\kern 0.18em\optbar{\kern -0.18em \PLambda}{}\xspace}

\def\AT#1     {\ensuremath{A_{\mathrm{T}}^{#1}}\xspace}           

\def\C#1      {\ensuremath{\mathcal{C}_{#1}}\xspace}                       
\def\Cp#1     {\ensuremath{\mathcal{C}_{#1}^{'}}\xspace}                    
\def\Ceff#1   {\ensuremath{\mathcal{C}_{#1}^{\mathrm{(eff)}}}\xspace}        
\def\Cpeff#1  {\ensuremath{\mathcal{C}_{#1}^{'\mathrm{(eff)}}}\xspace}       
\def\Ope#1    {\ensuremath{\mathcal{O}_{#1}}\xspace}                       
\def\Opep#1   {\ensuremath{\mathcal{O}_{#1}^{'}}\xspace}                    

\newcommand{\tev}{\ifthenelse{\boolean{inbibliography}}{\ensuremath{~T\kern -0.05em eV}\xspace}{\ensuremath{\mathrm{\,Te\kern -0.1em V}}}\xspace}
\newcommand{\gev}{\ensuremath{\mathrm{\,Ge\kern -0.1em V}}\xspace}
\newcommand{\mev}{\ensuremath{\mathrm{\,Me\kern -0.1em V}}\xspace}
\newcommand{\kev}{\ensuremath{\mathrm{\,ke\kern -0.1em V}}\xspace}
\newcommand{\ev}{\ensuremath{\mathrm{\,e\kern -0.1em V}}\xspace}
\newcommand{\gevc}{\ensuremath{{\mathrm{\,Ge\kern -0.1em V\!/}c}}\xspace}
\newcommand{\mevc}{\ensuremath{{\mathrm{\,Me\kern -0.1em V\!/}c}}\xspace}
\newcommand{\gevcc}{\ensuremath{{\mathrm{\,Ge\kern -0.1em V\!/}c^2}}\xspace}
\newcommand{\gevgevcccc}{\ensuremath{{\mathrm{\,Ge\kern -0.1em V^2\!/}c^4}}\xspace}
\newcommand{\mevcc}{\ensuremath{{\mathrm{\,Me\kern -0.1em V\!/}c^2}}\xspace}

\def\cm   {\ensuremath{\mathrm{ \,cm}}\xspace}
\def\cma  {\ensuremath{{\mathrm{ \,cm}}^2}\xspace}
\def\mm   {\ensuremath{\mathrm{ \,mm}}\xspace}

\def\mum  {\ensuremath{{\,\upmu\mathrm{m}}}\xspace}

\def\sec  {\ensuremath{\mathrm{{\,s}}}\xspace}

\def\degc {\ensuremath{^\circ}{C}\xspace}

\def\bar   {\ensuremath{{\mathrm{ \,bar}}}\xspace}
\def\W   {\ensuremath{{\mathrm{ \,W}}}\xspace}
\def\kW   {\ensuremath{{\mathrm{ \,kW}}}\xspace}

\def\gsim{{~\raise.15em\hbox{$>$}\kern-.85em
          \lower.35em\hbox{$\sim$}~}\xspace}
\def\lsim{{~\raise.15em\hbox{$<$}\kern-.85em
          \lower.35em\hbox{$\sim$}~}\xspace}

\def\tell1  {TELL1\xspace}
\def\ukl1   {UKL1\xspace}

\def\cotwo         {\ensuremath{\mathrm{ CO_2}}\xspace}

\def\co2   {\ensuremath{\mathrm{ \,mm}}\xspace}

%% file: 01_introduction.tex
\section{Introduction}

The LHCb collaboration is replacing its VErtex LOcator (VELO) detector 
during Long Shutdown 2 of the LHC as part of a major Upgrade programme to allow the experiment to operate at 
an instantaneous luminosity of $\mathcal{L} = 2 \times 10^{33}~\cm^{-2}\sec^{-1}$, 
five times higher than that of previous runs~\cite{LHCb-TDR-013}. 
The VELO requires very precise tracking and fast pattern recognition
in order to reconstruct collisions and decay vertices in real
time as the first step of the LHCb trigger decision.  A new hybrid pixel detector replaces the original silicon strip detector~\cite{jinstdetectorpaper,VELOTDR}, equipped with a lightweight and highly thermally efficient cooling solution of evaporative $\rm{CO_2}$ circulating in microchannels embedded in a silicon cooler.  

\begin{figure}[ht]
 \centering
     \includegraphics[width=0.99\textwidth]{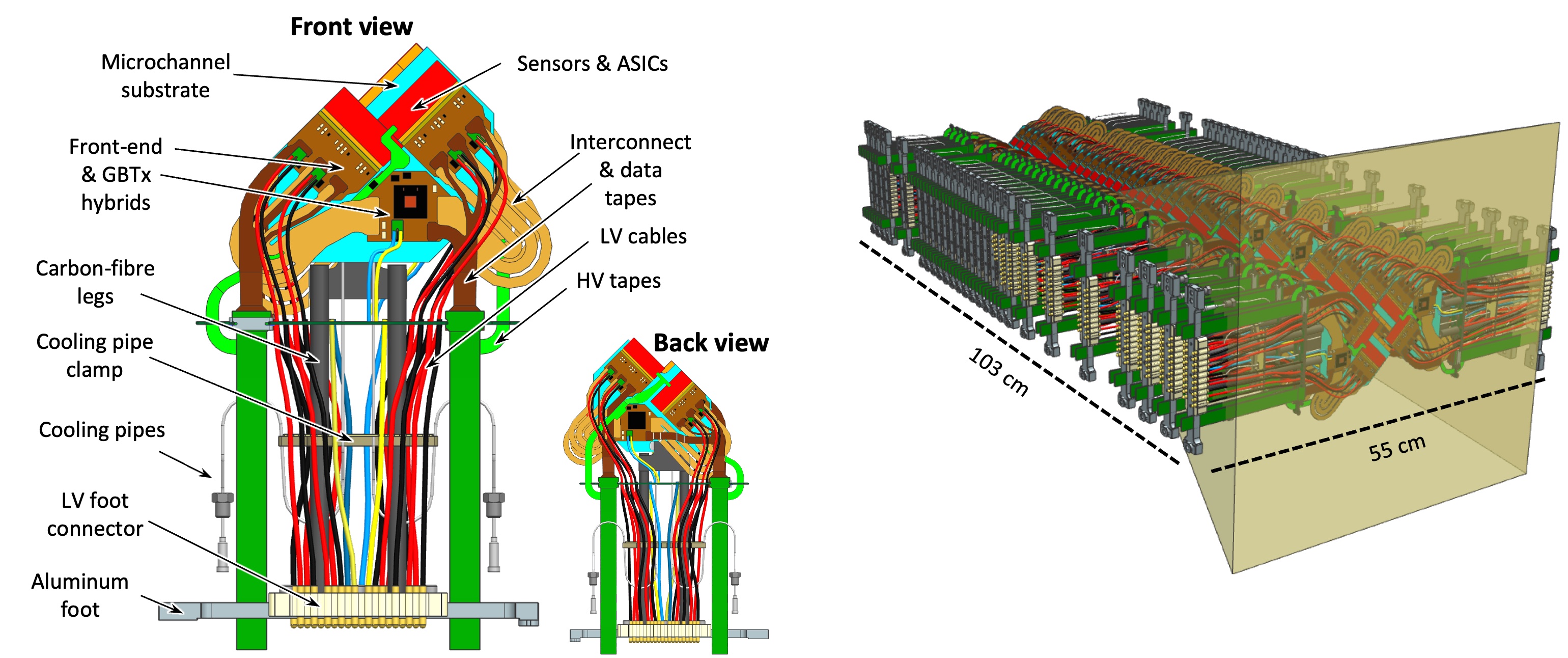}
    \caption[]{Diagrammatic view of the module layout (left), highlighting the hybrid pixel sensors (two on each side) and the optical link chip (GBTx) hybrid (one on each side), which are the main power-dissipating elements.  Module layout in LHCb (right).  The shaded region indicates the LHCb acceptance, showing that the inactive areas of the downstream modules fall in the acceptance.}
  \label{fig:module_front_back_cross_section}
\end{figure}

The sensing elements of each module consist of four planar silicon sensors, each bump-bonded to three VeloPix~\cite{VELOPIX} readout ASICs.   A diagrammatic view of the module and the arrangement of the two VELO halves is shown in \cref{fig:module_front_back_cross_section}.   The sensors are approximately $43.5 \times 15 \mm^2$ in size and are arranged with two on each side of the module.  The principal sources of heat on the module are the VeloPix ASICs which can dissipate up to $1.9 \W$ ($\sim$1$\W/\cm^2$) each depending on the luminosity and the operation mode, the sensors, which may dissipate about $1 \W$ per tile after irradiation (for an integrated luminosity of 50$\rm{fb^{-1}}$), and the GBTx hybrids \cite{karol_2681428}, one on each side of the module, which dissipate approximately $1.6 \W$ each.  The total nominal power budget per module is conservatively $ 30 \W$, giving a total of $1.56 \kW$ for the 52 modules of the whole VELO.  The heat is dissipated in vacuum, requiring a suitably efficient cooling system.  The sensors will experience an integrated irradiation of up to $\phi =$ \SI{8e15}{\mevneq}, with a highly non-uniform distribution concentrated at the tip of the sensor.  The sensor tips need to be kept at -20\degc or below in order to reduce reverse current and reverse annealing, while being directly bump bonded to the heat generating ASICs, placing stringent requirements on the cooling.  At the same time the material budget must be kept to a minimum, in order to allow accurate VELO reconstruction of low momentum tracks originating from heavy flavour decays and to be able to identify displaced tracks and vertices.  This is particularly true at the tip of the module, which is important for the extrapolation to the original primary or secondary vertex, and for this reason the cooling substrate is retracted from the tip to minimise the material.  It is also true over the full module height, due to the fact that the downstream modules are entirely within the LHCb acceptance~\cite{LHCb-TDR-013}.  The cooling system must be capable of continuous operation such that the sensors can be kept cool to suppress reverse annealing even during long shut down periods of the LHC.  Finally, the VELO system is enclosed within the secondary vacuum of the LHC and will be almost completely inaccessible for the decade of scheduled running time, hence the reliability of the system is paramount.

Based on the successful operation of the \cotwo cooling based on a 2-phase accumulator controlled loop~\cite{bart_original,bart_paolo_2pacl}  implemented for LHCb Run 1, this principle is maintained for the cooling plant of the VELO Upgrade.  Two identical \cotwo cooling plants, which share a common chiller, have been built, capable of simultaneously removing $1.6 \kW$ of heat from the VELO and $5.4 \kW$ of heat from the silicon upstream tracker at $-35$ \degc~\cite{jerome_mauve} for the set temperature at the cooling plant accumulator. As for the previous VELO an evaporator system~\cite{Verlaatevaporator:2010nl} is designed to distribute the coolant to the modules, however the previous solution for the thermal interface of cooling pipes embedded in Aluminium ``cookie'' heat spreaders~\cite{VanLysebettenlhcbrun1:2008vu} does not satisfy the pixel upgrade requirements and a new solution had to be implemented.

\subsection{Microchannel cooling overview}

The idea of integrating coolant in microscopic channels was first proposed~\cite{water_uchan} as a method of achieving high performance cooling of planar integrated circuits, with water as the coolant.  For laminar flow in confined channels, the heat transfer coefficient scales inversely with the channel width, ultimately limited by the coolant viscosity.  The microchannel concept with channels integrated directly into a silicon wafer is very attractive for active cooling of planar silicon components for many reasons above the high thermal efficiency provided.  It benefits from no coefficient of thermal expansion mismatch between the substrate and the silicon components to be cooled, which would otherwise have to be absorbed by an intermediate glue layer. In this case the silicon components are the hybrid pixel tiles, which are the major source of heat on the module and also the most critical in terms of temperature; the other active components are mounted on a flex circuit with a flexible glue interface.  The radiation length is naturally low due to the small size of the channels and the material distribution is very homogeneous compared to a solution incorporating bulky pipes. The silicon itself has a high thermal conductivity; $\sim$150~$\rm{W/(m \cdot K)}$ at room temperature, improving to $\sim$190~$\rm{W/(m \cdot K)}$ at a typical operational temperature of -25\degc~\cite{ThermalCondSi}. By using an appropriate layout of the channels, the cooling can be routed precisely to the areas where it is needed. The natural planar format is an excellent match to the planar components to be cooled, maximising the geometric thermal contact.  Using standard silicon processing techniques additional features are available, such as metallisation of alignment marks and solder footprints\footnote{althugh not used for LHCb, features such as metallic traces may be used in the future} and the wafers can be diced to produce final coolers with complex shapes.  

\subsection{Microchannel cooling for the LHCb VELO and synergy with parallel R\&D efforts}

\begin{figure}[ht]
 \centering
     \includegraphics[width=0.99\textwidth]{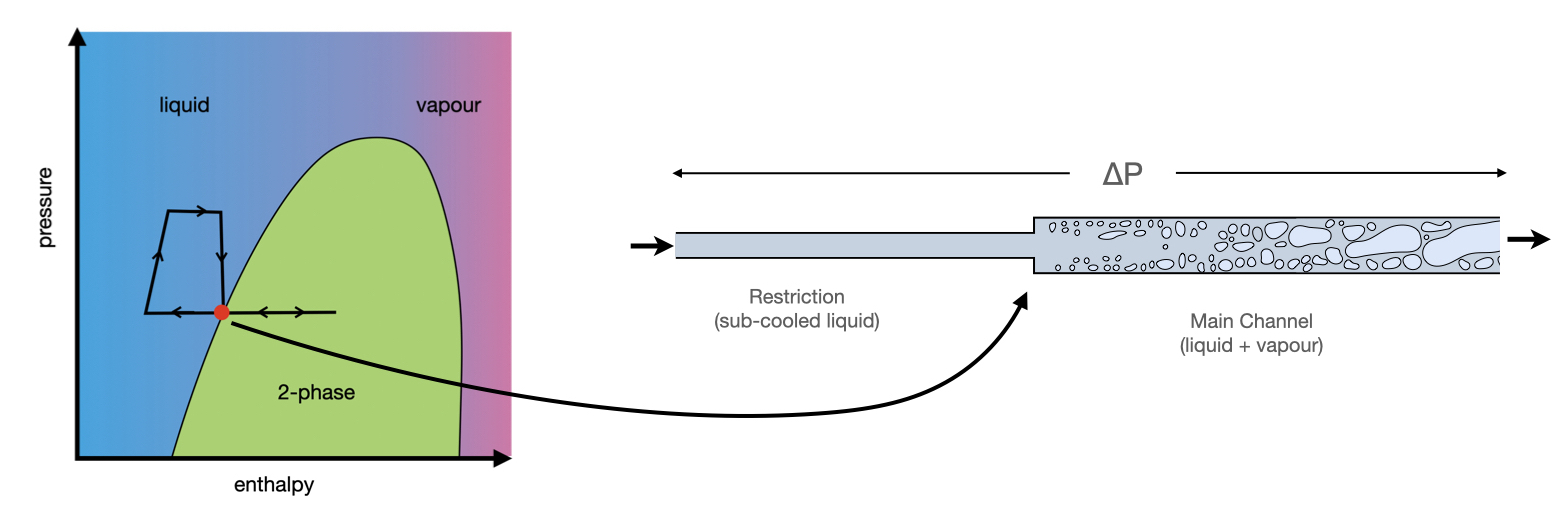}
    \caption[]{Typical channel shape for the bi-phase \cotwo microchannel cooling implementation.  The pressure drop at the point where the channel expands should bring the coolant to the saturation point as it enters the region of the detector to be cooled.  The diagram on the left illustrates the principle of the Two-Phase Accumulator Controlled Loop (2PACL) cooling concept used in LHCb~\cite{bart_original}.}
  \label{fig:channel_shape}
\end{figure}

For the LHCb VELO implementation, the choice has been made to use microchannel cooling in combination with bi-phase \cotwo as the coolant.  \cotwo is ideal as a coolant due to its high latent heat capacity and low viscocity, and it is radiation hard.  Bi-phase cooling, where the heat from the electronics is absorbed by the evaporating liquid \cotwo, gives powerful cooling capability and is stable against changes in power consumption of the module during running. As illustrated in \cref{fig:channel_shape}, the \cotwo enters the microchannel structure as sub-cooled liquid through a restriction, before entering the main channel where it starts to boil.  A fraction of the liquid is then evaporated while flowing under the heat sources.  The VELO modules are designed for a typical flow of 0.4 $\rm{g/s}$, which for 30W heat dissipation would give a final vapour quality of approximately 25\%, which gives a good thermal efficiency while still remaining safely away from `dry-out' conditions, where the \cotwo completely evaporates before the end of the channel.  The restriction dominates the fluidic resistance and so provides excellent control, ensuring even flow in the main channels and avoiding instabilities.  The coolant temperature decreases from the input to the output of the module due to the bi-phase nature of the cooling associated with the variation of pressure to keep the coolant flow.  A major challenge of the system relates to the operational pressures; for normal operation at -30\degc the system is at $\sim$14 \bar, while at room temperature (20\degc) the pressure reaches approximately $57 \bar$.  For safety reasons the entire system must be validated to $186 \bar$.

\begin{figure}[ht]
 \centering
     \includegraphics[width=0.99\textwidth]{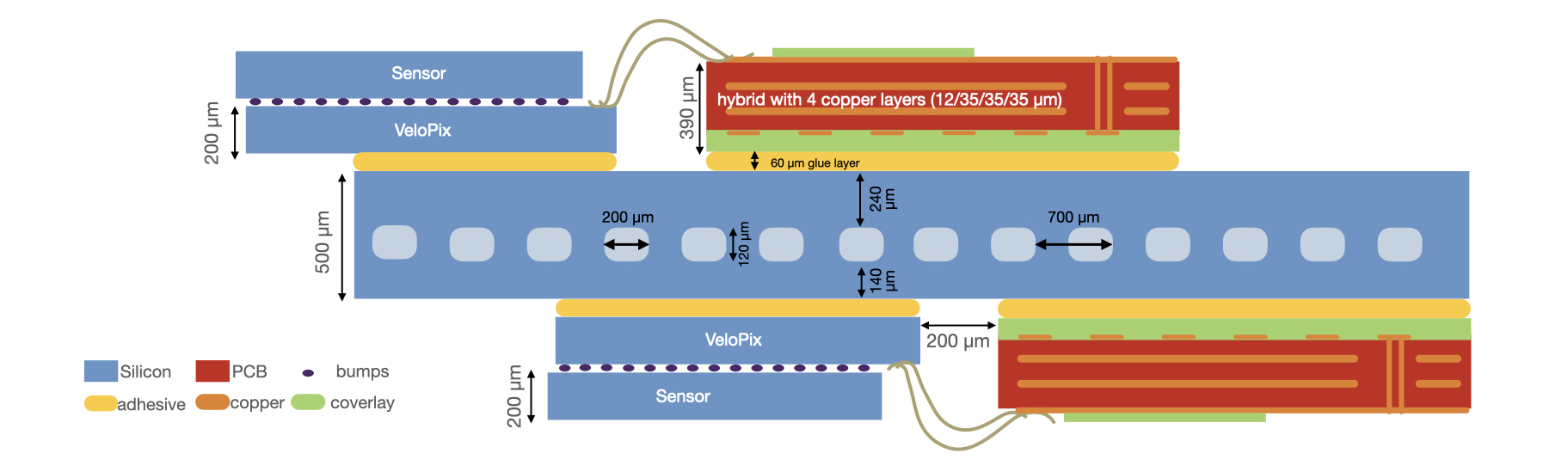}
    \caption[]{Cross section of the VELO module (not to scale).  The sensors overhang the cooler at the tip of the module by 5 \mm in order to keep the material budget to the minimum for the first measured points on each track.}
  \label{fig:module_cross_section}
\end{figure}

A cross section of the microchannel structure as implemented in the VELO modules is shown in \cref{fig:module_cross_section}.  By using the microchannel layout a very homogeneous material distribution can be achieved and the coolant can be delivered directly under the heat sources. There is a distance of just $140/240 \mum$ between the coolant and the surface of the microchannel cooler.  The thermal barrier is dominated by the glue layer, which has a typical thickness of $60 \mum$, and low thermal gradients can be achieved across the module.  The cooling is found to be effective enough that it is possible to withdraw the microchannel 5\mm from the tip of the module, in order to keep the material budget to the minimum for the first measured points on each track.  

As this was an emergent technology at the time of the adoption into LHCb, a long R\&D path was necessary to overcome the main challenges.  At the start a suitable facility had to be found in order to be able to make prototypes and demonstrate the basic principles.  Following this a manufacturer had to be identified capable of producing microchannel coolers with such large dimensions with high enough quality to satisfy the pressure requirements.  The R\&D also had a large focus on the method of attaching the fluidic connector, and establishing suitable quality control processes to ensure long term operational stability.

The work described in this paper was carried out by the LHCb VELO collaborating institutes, and also benefitted from a high degree of synergy with other experiments, in particular the NA62 GigaTracker~\cite{AglieriRinella:2019eri}, which implements a pioneering microfluidic cooling system with a different coolant and with different requirements, but has many specifications in common.  At CERN the R\&D was supported by the EP-DT group and the collaboration set up between CERN and EPFL which allowed quick prototyping at the MicroNano Technology centre.  The EP-DT group is continuing to conduct research into questions related to an effective and reliable use of micro-structured thermal management devices for HEP detectors, working in the framework of the CERN strategic R\&D programme on technologies for future experiments~\cite{Aglieri:2764386}.


%% file: 02_prototyping.tex
\section{Microchannel technology prototyping and design optimisation}
\label{sec:prototyping}

The principle of microchannel manufacture is to etch trenches into the surface of a silicon wafer then atomic bond a so called `cap' wafer to close the channels.  Suitable exit and entry holes are incorporated, typically by etching after bonding, such that liquid can circulate through the capillaries.  To fabricate such a system many steps are involved, including photolithography, etching, chemical mechanical polishing, bonding and dicing.  The substrates may be silicon,
plastic, or glass. For the VELO upgrade application both wafers in the final module are silicon.  However glass of various thicknesses has also been used as a cap wafer for several of the R\&D steps, since samples can be produced by the more accessible anodic bonding process, avoiding the need to go to industry\footnote{The samples used for the VELO R\&D steps were produced at the  MicroNano Technology centre at EPFL, Lausanne, Switzerland.}, 
and have the advantage of allowing a clear
view of the channels. 

\subsection{Proof of principle - Snake I}

In order to test the microchannel geometry and measure the heat and flow characteristics, silicon-pyrex bonding was used, because of the much more ready availability of anodic bonding technology.  In such a case the heat load is applied on the silicon side only, due to the much poorer thermal conductivity of the glass.  This method has the advantage that the channels can be directly observed through the glass. This prototype was successfully tested with circulating \cotwo supplied by a mobile cooling plant which provided liquid \cotwo at -40\degc with a pressure differential between the inlet and
outlet of $\Delta$P$\sim$4 \bar.  It was demonstrated that, with a $\Delta$P$\sim$3 \bar and a \cotwo
mass-flow of 0.15~g/s, $ 12.9 \W$ of power could be removed from the
module before the \cotwo enters the
`dry-out' condition.    The total heat removal was limited by the pressure difference which could be supplied by the cooling plant.  At maximum heat dissipation, the temperature differential
between the heater and the output was only 3\degc. From this the
thermal figure of merit (TFM) of the system can be determined
to quantify the effectiveness of the cooling design.  The
measured value was
\begin{equation}
  {\rm{TFM} = \frac{ \Delta T_{\rm heater - output} } 
                       { {\rm Power/\cma} } 
               = 1.5\ {\rm K\cma W^{-1}}}
\end{equation}
which demonstrates the high cooling efficiency potential of the microchannel design.  For comparison, conventional designs with cooling ledges can typically achieve a TFM of $ 20 \rm{~K~cm^2 W^{-1}}$, while highly integrated pipe-structure designs aim to reach values of TFM $\sim 12 \rm{~K~cm^2 W^{-1}}$~\cite{bart_paolo_2pacl}. This work was published
in \cite{Nomerotski:2012cn}.

\subsection{Test structures}

After the successful proof of principle a second round of prototyping was undertaken with multiple objectives.  A large number of silicon-silicon test samples were produced, to investigate the impact of design choices on the rupture resistance, to qualify different silicon bonding processes, and to test small scale circulating samples. In addition, silicon-glass structures were produced to investigate the resistance of the silicon cover thickness to rupture and the dependence on the channel width.

\begin{figure}[ht]
 \centering
     \includegraphics[width=0.99\textwidth]{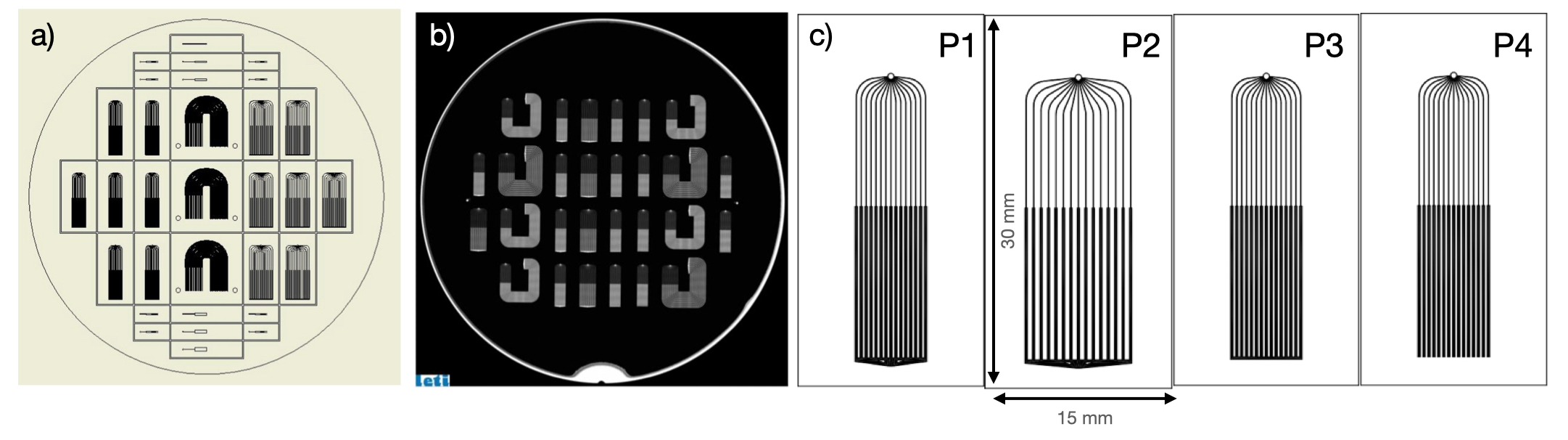}
    \caption[]{a) Silicon microchannel six-inch test wafer design produced at Southampton University Nanofabrication Optoelectronics  b) Scanning Acoustic Microscope image of microchannel eight-inch wafer produced at CEA-LETI Grenoble c) Detail of the P1,P2,P3,P4 test structures on the CEA-LETI wafer.}
  \label{fig:test_structures}
\end{figure}

The silicon wafer layouts tested are illustrated in \cref{fig:test_structures}.  For the tested samples all channels were etched to a depth of $70 \mum$ with a width of $200 \mum$ for the main channels and $30 \mum$ for the inlet restrictions. The total thickness of the samples was $400 \mum$. The \cotwo is supplied through an inlet hole with diameter $1.6 \mm$ to supply all the channels, and the samples have different sizes for the outlet manifold, with P4 having no outlet manifold at all.  Sample sets from two different bonding processes were received, {\em hydrophilic} and
{\em hydrophobic} which differ in the complexity of the surface
preparation.

\begin{itemize}

  \item The hydrophilic process is an industry-wide technique that
  introduces water molecules across prepared silicon surfaces resulting in the formation of
  Si$-$OH hydrogen bonds.  Upon contact of the surfaces the
  Si$-$OH groups readily bind to each other forming Si$-$O$-$Si
  covalent bonds.  These bonds increase in number and strength during
  an annealing (heating to 800\degc) during which the residual H$_2$O
  disperses in the crystal lattice.

  \item A hydrophobic process removes any oxide layer and allows
  direct Si$-$Si covalent bonding.  The disadvantage is that the
  surface quality must be of a much higher standard and prepared in a higher-specification clean room, to reduce the probability that voids (bubbles in the Si$-$Si bond) can form.  The potential advantage is a near-perfect
  homogeneous union of the silicon lattice.

\end{itemize}

\begin{figure}[ht]
 \centering
     \includegraphics[width=0.99\textwidth]{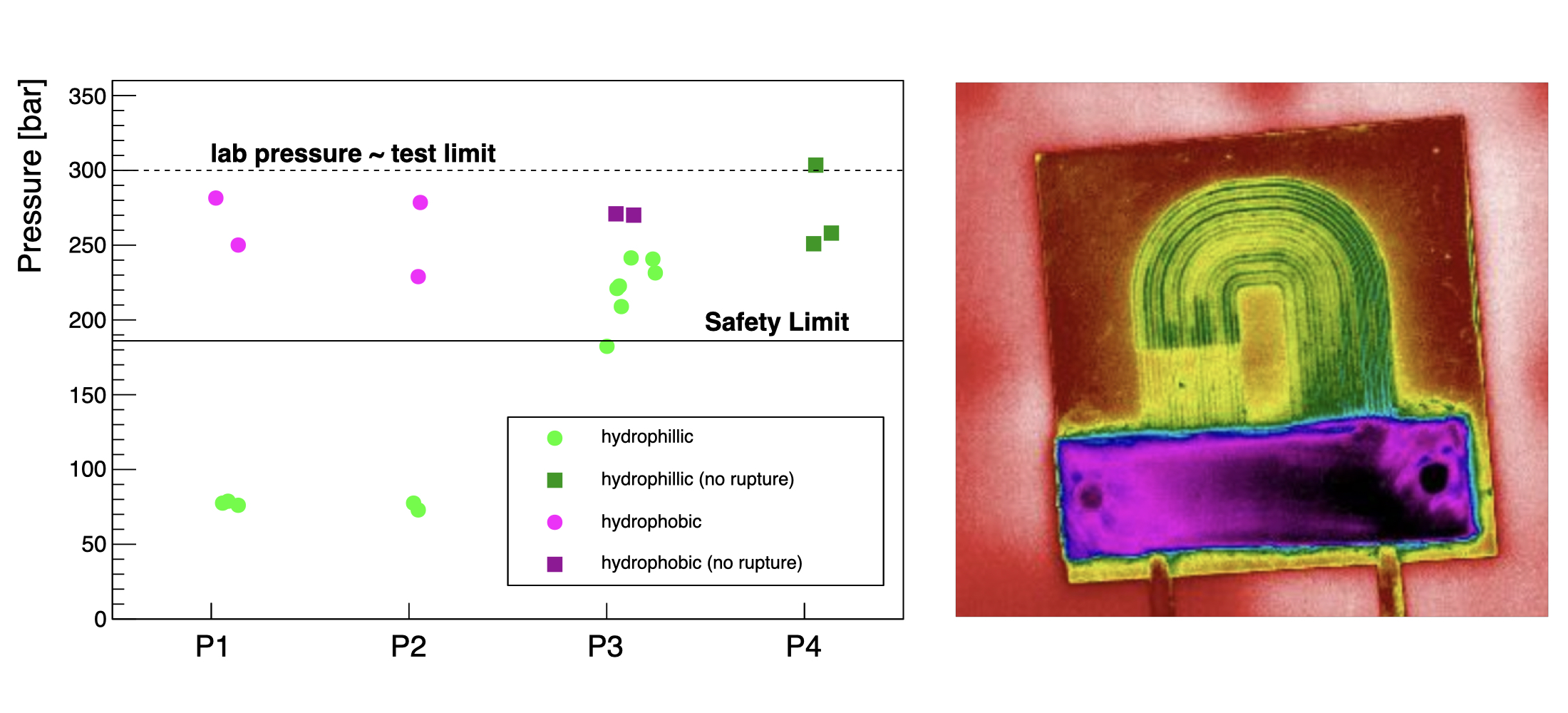}
    \caption[]{(left) Rupture tests of samples P1 to P4.  The P4 samples surpass the safety test for both hydrophillic and hydrophobic bonding.  (right) Infra-red camera image of water circulating in silicon-silicon bonded sample.}
  \label{fig:test_sample_results}
\end{figure}

Pressure tests were performed with water on the P1-P4 test structures.  It was possible to provoke breakages, but these all occurred at the large output manifold, not within the channels. The breakages seen in the hydrophilic bonding samples occurred both in the fusion layer and in the lattice, whereas the breakages from the hydrophobic bonded samples occurred only in the silicon lattice~\cite{LHCb-TDR-013}.  For the P4 samples it was not possible to provoke a breakage when testing up to $300 \bar$, which was the available experimental limit at the time this test was performed.  The results are shown in \cref{fig:test_sample_results}. To permit tests to higher pressure, silicon-silicon samples were clamped to reinforce the fluidic connection such that the integrity of the microchannels could be pressure-tested in isolation. Hydrophilic samples rupture in the body of the microchannel network at around $400 \bar$. The hydrophobic samples withstood the maximum pressure of the pump, which was $700 \bar$.

\begin{figure}[htb]
 \centering
     \includegraphics[width=0.95\textwidth]{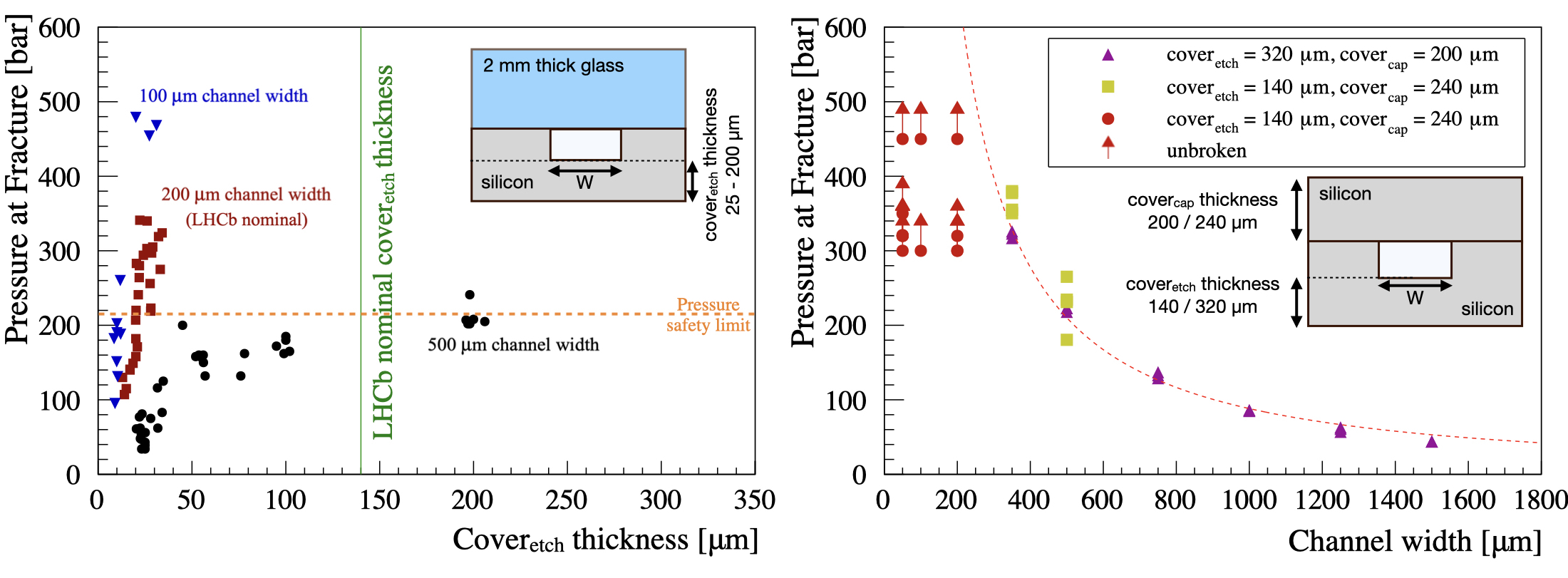}
    \caption[]{Results of investigations to determine dependence of rupture pressure on design parameters.  Left: Dependence on $\rm{cover_{etch}}$ thickness. It can be seen that the LHCb $\rm{cover_{etch}}$ thickness of $140 \mum$ is expected to be very resistant for the nominal channel width of $200 \mum$. Right: Dependence on channel width, for $\rm{cover_{etch}}$ and $\rm{cover_{cap}}$ thicknesses close to the LHCb nominal.  The line is an empirical fit to the data.  The red points with arrows did not break up to the pressure shown.  For this data set, those samples which broke all broke in the cap wafer.  Based on these data a maximum channel width of $200 \mum$ is set.}
  \label{fig:cover_thickness_and_width}
\end{figure}

Dedicated silicon-glass samples were used to investigate the dependence of the pressure resistance on the thickness of the silicon covering the cooling channel.  When measuring the thickness between the coolant and the wafer edge in the etched wafer, this is referred to as $\rm{cover_{etch}}$, and when measuring the corresponding thickness in the cap wafer, this is referred to as $\rm{cover_{cap}}$.  Rupture tests were performed on a number of samples of different silicon thickness and microchannel width. A clear relationship between rupture pressure and  $\rm{cover_{etch}}$ thickness is
observed for $500 \mum$ wide microchannels. The $200 \mum$ wide
microchannels, which is the nominal width for LHCb, tolerate pressures of more than $200 \bar$ even with a 30\mum $\rm{cover_{etch}}$ thickness.  For the nominal LHCb design thickness of 140\mum the rupture pressure obtained by extrapolation is beyond the scale of the plot, indicating that the fact that all hydrophobic samples hold $700 \bar$ is in line with expectation for the given $\rm{cover_{etch}}$ thickness.  A further set of samples were used to investigate the rupture pressure as a function of channel width for $\rm{cover_{etch}}$ and $\rm {cover_{cap}}$ thicknesses corresponding to the nominal $140 \mum$ and higher.  The rupture pressures rise to high values for lower channel widths, driving the choice of a maximum channel width of $200 \mum$ for the LHCb implementation as being a safe limit.  These results are illustrated in \cref{fig:cover_thickness_and_width}.

\subsection{Design optimisation - Snake II}

\begin{figure}[htb]
 \centering
     \includegraphics[width=1.0\textwidth]{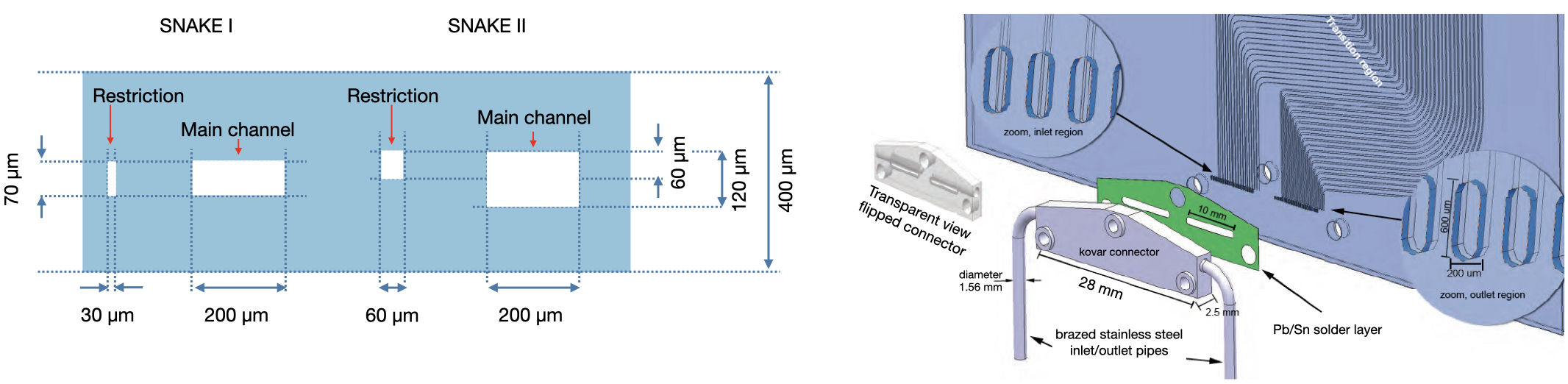}
    \caption[]{Design optimisation following prototyping rounds.  Left: Optimisation of channel dimensions. Right: Optimisation of manifold and connector region.}
  \label{fig:design_change}
\end{figure}

Following the first round of prototyping a second large scale prototype, dubbed ‘‘Snake II", was produced, with the aim of implementing and testing the major design changes before the manufacturing step.  The first change was to improve by a factor four the fluidic resistance without reducing the rupture resistance. This goal was achieved by moving to a two step etching procedure; in the first step etching the restrictions and main channels, and in a second step, the main channels only.  The main channel width was maintained at $200 \mum$, but the depth was increased to $120 \mum$, and the restrictions were made square, with dimension $60 \times 60 \mum^2$, decreasing the resistance and reducing the chance of clogging.  The issue of sensitivity to rupture in the manifold was removed by moving the manifold inside the connector, such that no channel in the silicon has a width greater than $200 \mum$.  The \cotwo is therefore distributed within the connector, and each channel has its own inlet and outlet.  These design changes are illustrated in \cref{fig:design_change}.

\begin{figure}[htb]
 \centering
     \includegraphics[width=0.75\textwidth]{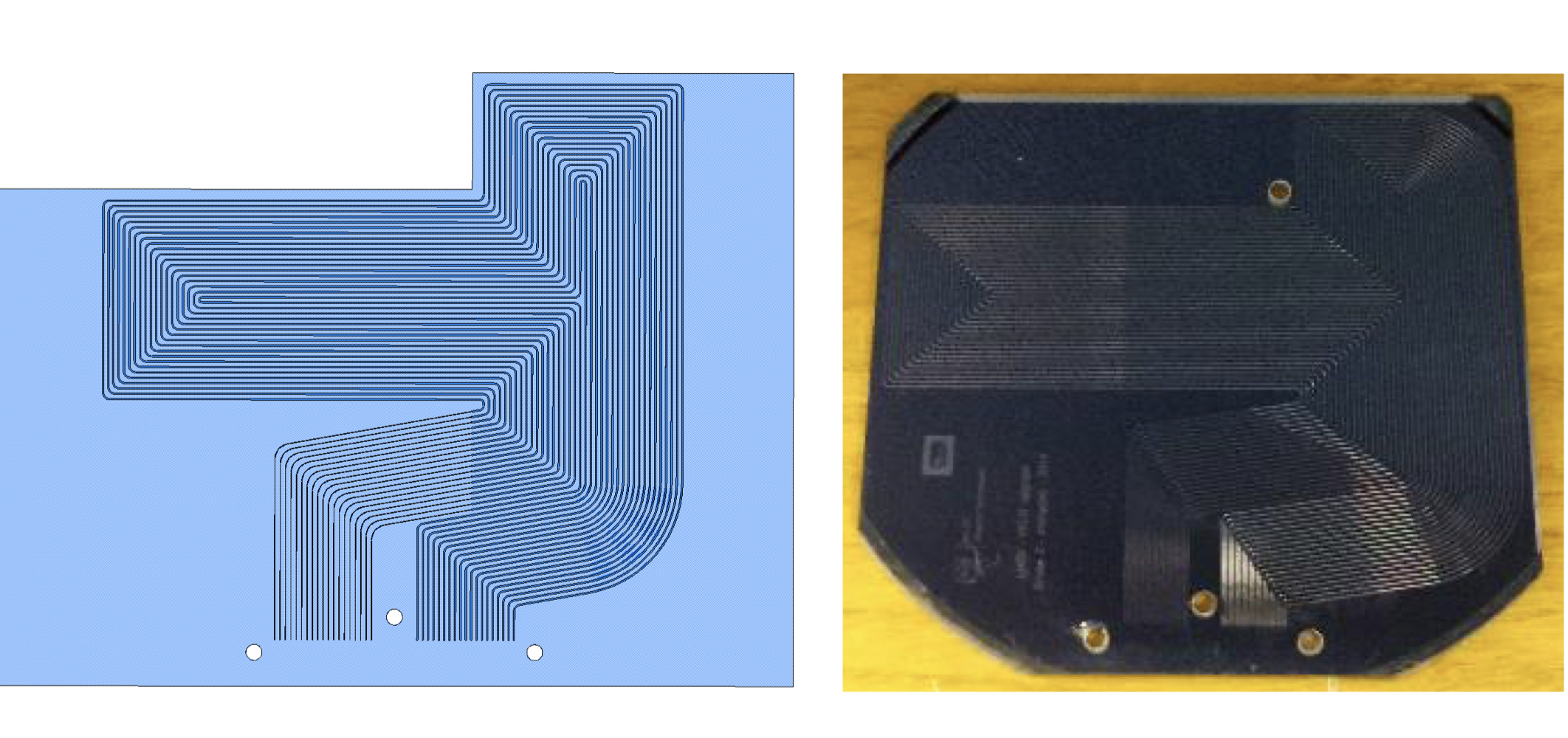}
    \caption[]{Snake II prototype design and manufacture,}
  \label{fig:snake2}
\end{figure}

The Snake II was successfully produced (see \cref{fig:snake2}) with  silicon-pyrex bonding technology. It featured 19 channels with similar length and fluidic resistance. It was equipped with thermal mockups made of $300 \mum$ thick silicon, metallised to simulate the heat density in the irradiated sensors, and thin film deposit heaters designed to mimic the ASIC heat distribution, all attached with conductive tape\footnote{Tape 3M 9461P}.  The heaters were all mounted on the silicon side of the Snake II.    As the test was designed to check the fluidic capabilities of the channels, a simple metallic clamp and O-ring was used for the connector and inlet/outlet cooling tubes.  The system was able to provide stable cooling up to $60 \W$ for temperatures between -25\degc and 15\degc set at the cooling plant and with \SI{60}{W} power dissipation giving a factor two safety margin.  At the lowest temperature the flow was approximately 0.3 g/s for a $\Delta$P of $4 \bar$.  The flow characteristic was found to be fairly well described, assuming that the pressure drop is dominated by the flow through the orifice, by $ F \sim \Delta{\rm P} \times C_v$.  The flow coefficient $C_v$ for Snake II was found to be approximately 2.1 g/s/$\sqrt(\Delta P)$, three to four times greater than that measured for Snake I.  These tests validated the design principles for the LHCb microchannels.

%% file: 03_LHCb_implementation.tex
\section{LHCb implementation}
\label{sec:implementation}

\subsection{LHCb microchannel design}

\begin{figure}[ht]
 \centering
     \includegraphics[width=0.99\textwidth]{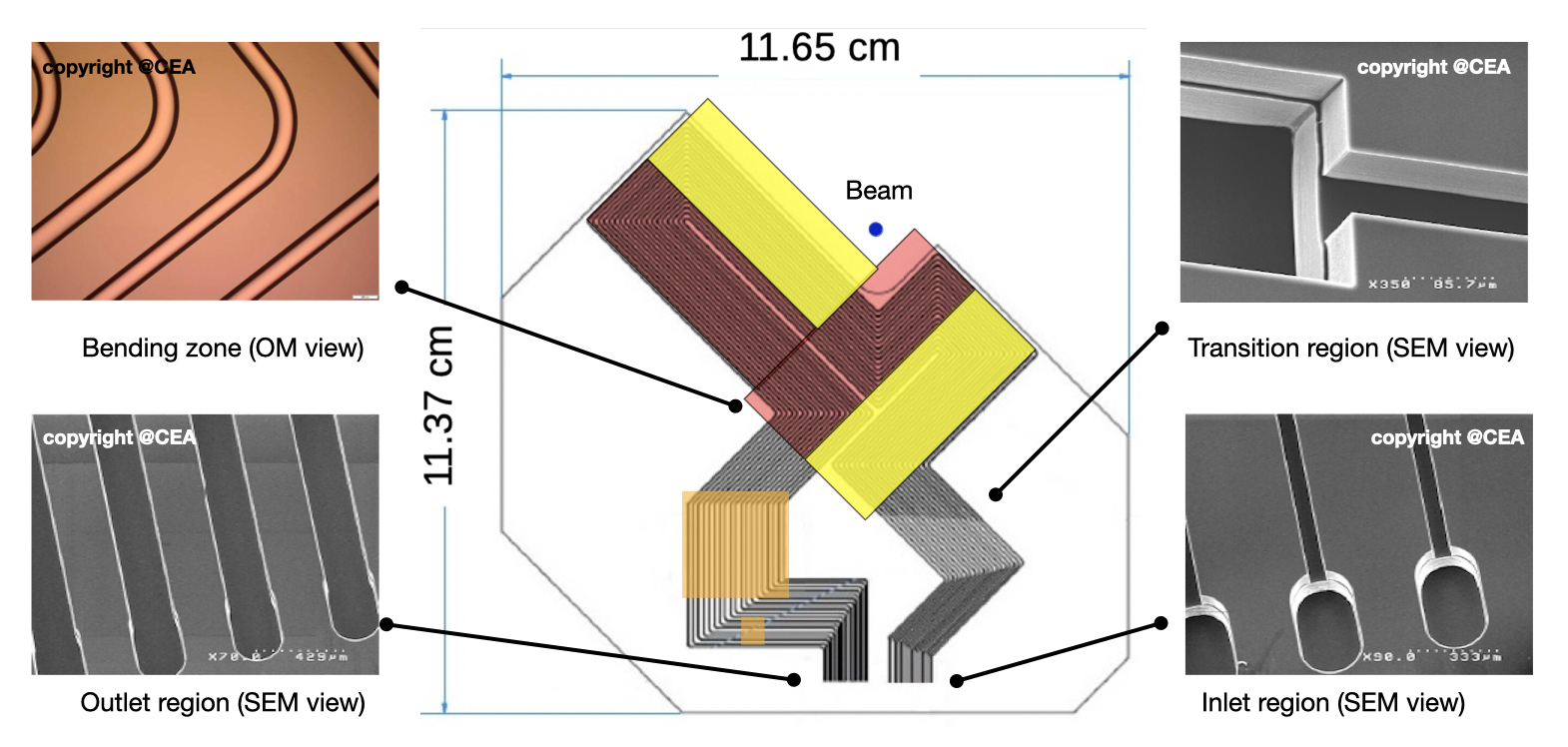}
    \caption[]{Layout of the final microchannel design, overlaid with the hybrid pixel tiles and heat producing ASICs (GBTx and the smaller GBLD).  The tiles on the front of the cooler are shown in yellow, those on the back in pink, and the position of the GBTx/GBLD chips on both sides of the module in orange.  Scanning electron microscope and optical microscope images of details from the production are also shown.}
  \label{fig:lhcb_design}
\end{figure}

The microchannel cooler design as implemented for LHCb and manufactured by CEA-LETI is illustrated in \cref{fig:lhcb_design}.  Viewed from the connector attachment side, the \cotwo enters as subcooled liquid on the right, passes through a $40 \mm$ long restriction, and ideally starts boiling in the main channels just before flowing under the first ASIC.  The channels are routed to pass under the heat dissipating hybrid pixel sensors and the GBTx hybrid.  There are a total of 19 channels and every bend is designed with the same radius ($0.5 \cm$) to ensure an equivalent pressure drop due to bending on all the channels.  The dimensions of the 19 main channels are $120 \times 200 \mum^2$, with a pitch varying between $450 \mum$ (at the outlet), $700 \mum$ (in the main body) and $990 \mum$ (under the GBTx hybrid).  The length of the main channels (after the restrictions) varies between 231 and 292 mm.  The inlet restrictions have dimensions of $60 \times 60 \mum^2$ and a pitch of $450 \mum$.  The total thickness of the wafer is $500 \mum$, and the channels are positioned asymmetrically in the final wafer, with $140 \mum$ silicon between the channel and the outside of the wafer on the connector side  ($\rm{cover_{etch}}$) and $200 \mum$ between the silicon and the outside of the wafer on the non-connector side ($\rm{cover_{cap}}$).  The inlet and exit holes are identical for all channels with dimensions of approximately $200 \times 600 \mum^2$.  The cooler has similar dimensions as Snake II but the positions where the sensors will be glued have been rotated by 45 degrees around the beamspot.  This decision, taken to ease the VELO insertion, however poses an additional challenge for the cooler production as the introduction of an internal angle to be cut out prevents saw dicing being used in the cooler production.

\begin{figure}[ht]
 \centering
     \includegraphics[width=0.99\textwidth]{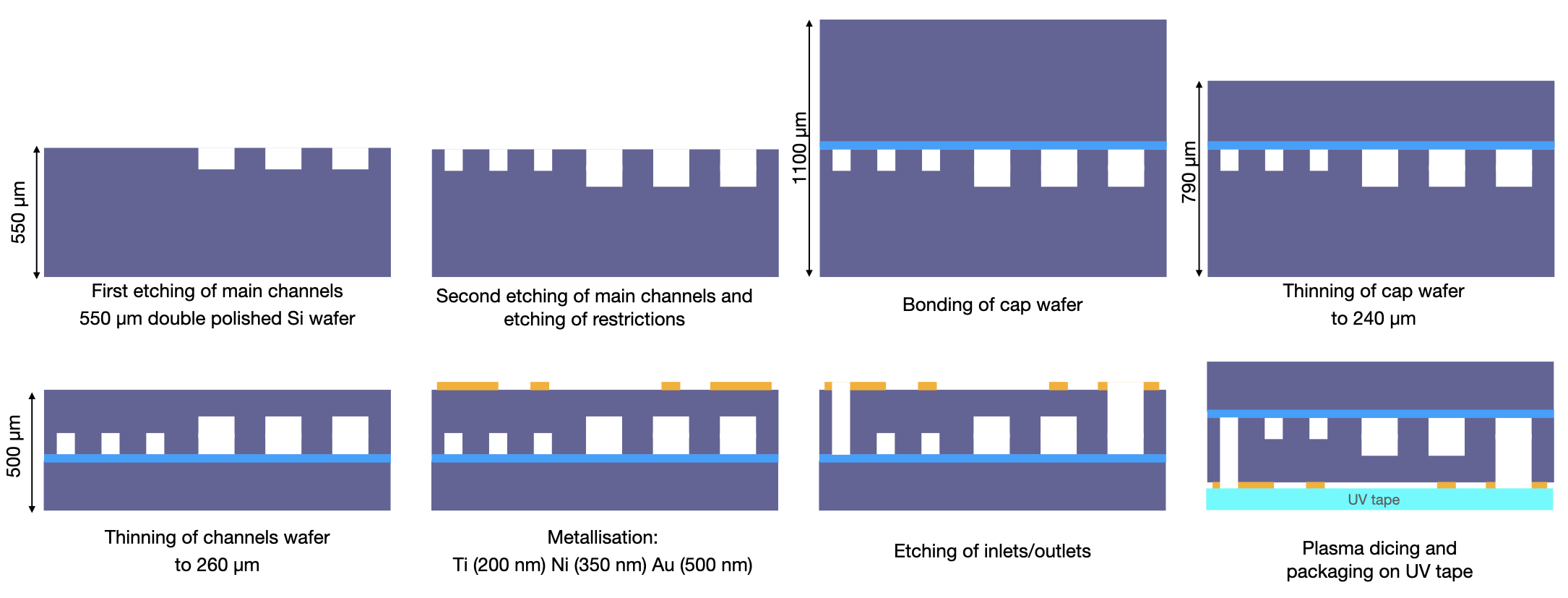}
    \caption[]{Schematic of the principal steps of the production process.}
  \label{fig:process_steps}
\end{figure}

\begin{figure}[ht]
 \centering
     \includegraphics[width=0.99\textwidth]{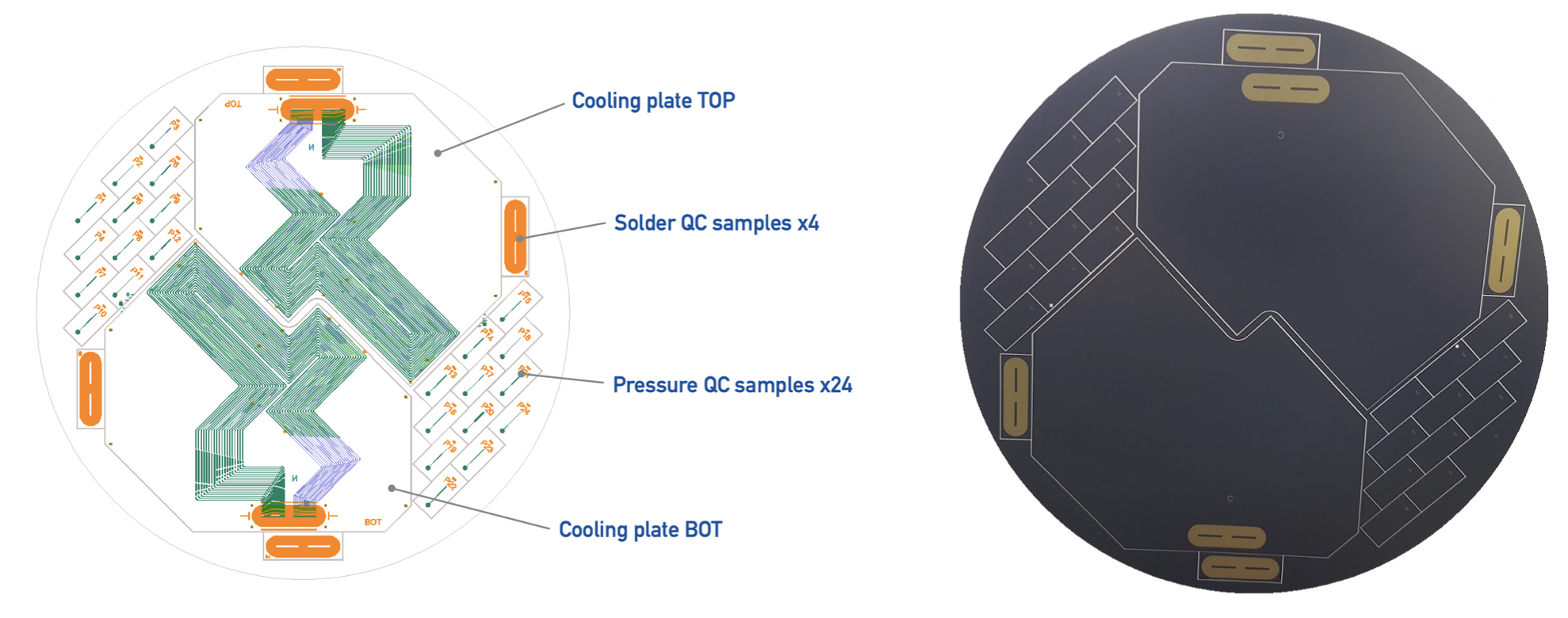}
    \caption[]{Left: Layout of the final wafer as manufactured.  There are two coolers, four additional connector footprints for metallisation and adhesion tests, and 24 pressure control samples.  Right: Photograph of the back of a processed and diced wafer.}
  \label{fig:wafer_and_photo}
\end{figure}

The principal process steps are illustrated in \cref{fig:process_steps}.  The starting material is double polished eight inch diameter silicon wafers with total thickness variation of $< 1 \mum$ and the thickness of the final device is $500 \mum$.  The channels are etched in the first wafer, before the bonding step with the cap wafer followed by an annealing process.  The cap wafer, then the channel (etched) wafer are thinned.  The metallisation, which provides alignment marks and the solder footprint for the connector is applied. (On the backside of the wafer, alignment marks are etched for plasma dicing and for reference marks during detector assembly.)  In the final step the cooler singulation is achieved by plasma dicing and the final wafer components are delivered on UV tape.  Two coolers can be fitted onto one eight inch diameter wafer. The layout of the final wafer is illustrated in \cref{fig:wafer_and_photo} along with a photo of the back face of a completed wafer.  The additional wafer space was filled with dedicated samples to check the metallisation quality for the connector soldering pad and small channels which could be used for pressure testing and bonding quality.

\subsection{Quality assurance}

The quality assurance at the production stage was based on the following criteria:

\begin{itemize}
    \item No defects in the high resolution ($<$ 50 \mum) scanning acoustic microscope (SAM) images taken after post-bonding annealing can be detected in the vicinity (within 500 \mum) of the channels, inlet or outlet areas.
    \item The edges of the devices do not present any abnormal defects caused by the dicing, such as cracks or chipping.
    \item The metallisation stack for soldering the fluidic connector has good surface quality and adherence to the silicon surface for subsequent soldering of the fluidic connector.  The surface quality, adherence to the silicon surface and wettability can be checked with the additional samples on the wafer.
    \item The pressure tests of the test structures on the same wafer show consistent pressure resistances with the measurements during the R$\&$D phase. 
\end{itemize}

\begin{figure}[ht]
 \centering
     \includegraphics[width=0.99\textwidth]{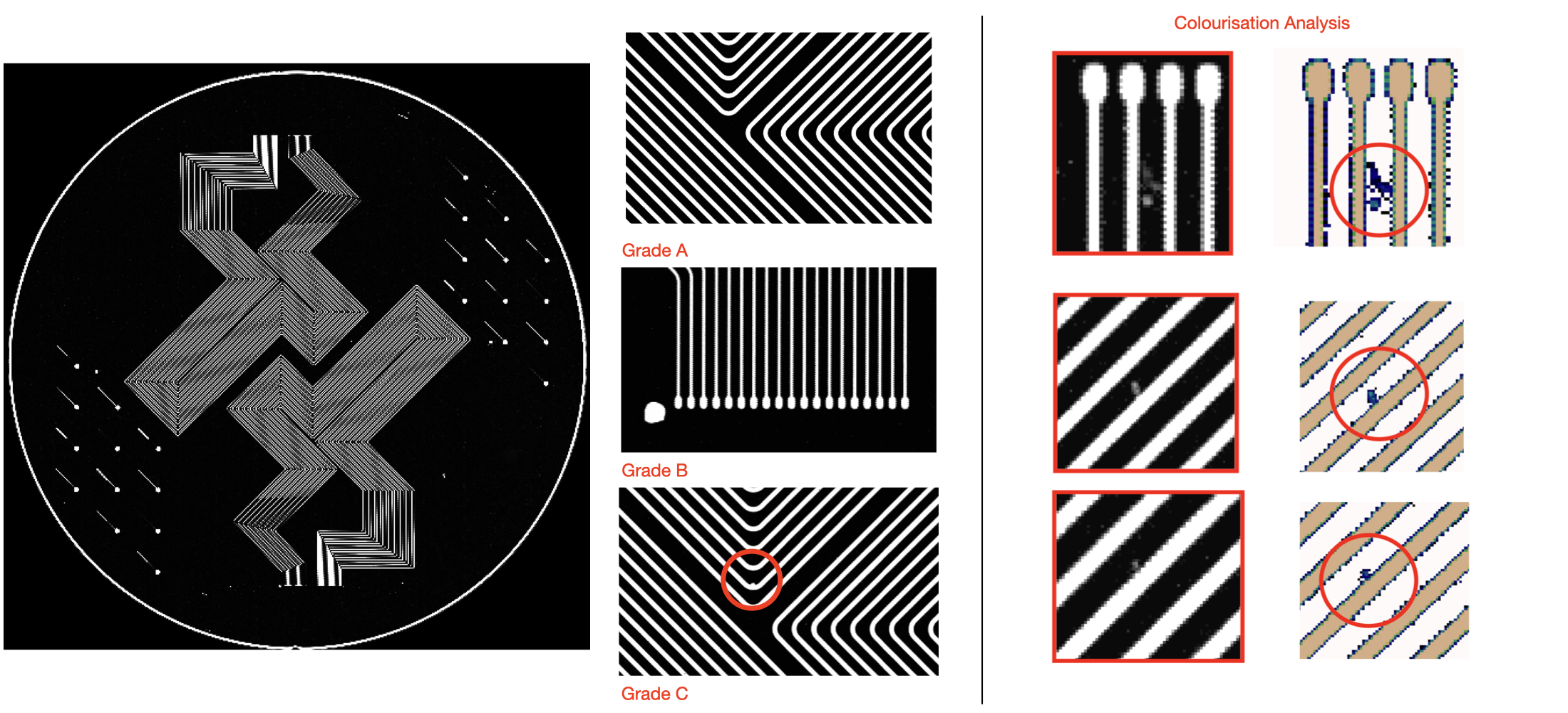}
    \caption[]{Left: Typical SAM image, with details showing examples of classifications into Grade A, B and C.  Right: illustration of regions with a defect together with the image after colourisation.}
  \label{fig:sam_analyses}
\end{figure}

These requirements were challenging, in particular the first one, which is based on the concept that no channel width should exceed $200 \mum$, as discussed in \cref{sec:prototyping} and supported by the measurements in \cref{fig:cover_thickness_and_width}.   The identification of defects from the SAM images required painstaking scanning by eye of multiple high-resolution images.  A grading procedure was developed, classing perfect coolers with no defects as A grade, coolers with defects in the input/output region as B grade and coolers with defects near channels as C grade.  Small dicing imperfections were accepted if the distance perpendicular to the ideal dicing line to the tip of the imperfection is below  300\mum and their location did not affect the module assembly procedure afterwards. As the connector region is reinforced with a piece of silicon glued on the back of the cooler, small defects in the inlet/outlet region could possibly be accepted but out of an abundance of caution these were not used for installation in the experiment. The coolers are large, with just two fitting onto one wafer.  For this reason the requirement that there should be no defect at all within the channel region of each cooler had a large impact on the yield during production.  Not taking into account wafers discarded during production and dicing losses, approximately 69\% of produced coolers were accepted for modules to be installed into LHCb.  Due to the difficulty of scanning the images by eye and the dependence on the operator, image processing software was produced to make any defects easier to identify.  The images were transformed to colour scale, and background noise was suppressed.  This made the defects easier to spot, as shown in \cref{fig:sam_analyses}.  All defects were reverified on the original SAM images before final classification.

\FloatBarrier

%% file: 04_performance_simulation.tex
\section{Performance simulation} 
\label{sec:simulation}

\subsection{FEA simulations}

\begin{figure}[htbp]
\begin{center}
  \includegraphics[width=0.8\textwidth]{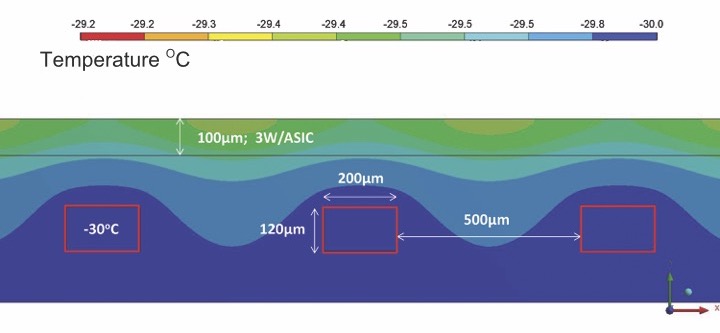}
  \caption{Thermal map from the microchannel FEA simulations, with a 
    cross section illustrating the optimal channel separation and the very
    small temperature gradients to the heat source.}
  \label{much_FEA0.5}
\end{center}
\end{figure}

\begin{figure}[htbp]
  \centering
  \includegraphics[width=0.8\textwidth]{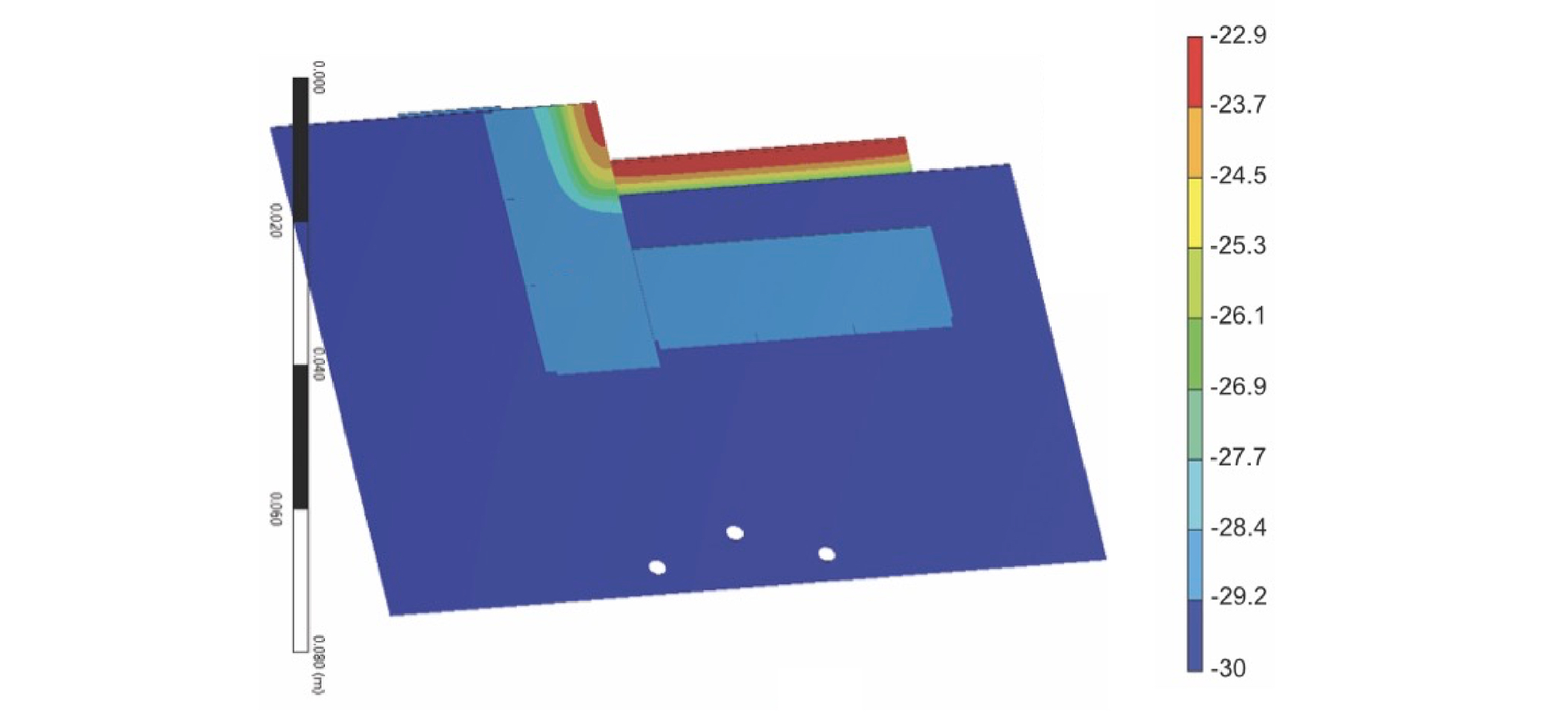}
  \caption{Thermal map from the microchannel FEA simulations, with a 
   view of
    the full module at nominal power. The maximum $\Delta$T to
    the silicon tip, from which the cooler is retracted, is
  estimated to be about 7\degc.}
  \label{much_FEA1}
\end{figure}

A finite element analysis was used to quantify the heat flow
through a microchannel cooler and to optimise the spacing between the
microchannels. This study modelled a homogeneous $400 \mum$ silicon
cooler containing $70\mum\times200\mum$ microchannels. A thin heat
source on one side provides $ 3 \W$ / ASIC ($\sim 36 \W$ per module), comparable to the power dissipation achieved in the final modules.  The study indicates that a separation of $500 \mum$ between channel edges is optimal, allowing efficient heat exchange along three sides of the channel, instead of being dominated by the side closest to the heat source which is the case for smaller pitches.  In addition the larger spacing provides maximal bonding surface.  With this design the simulation shows a negligible temperature differential between the coolant and heated surface, as shown in \cref{much_FEA0.5}.
The full module under maximum power with a coolant temperature of
-30\degc shows an expected $\Delta$T to the silicon tip of about
7\degc, as illustrated in \cref{much_FEA1}.

\FloatBarrier

\subsection{CoBra simulations}

\begin{figure}[htbp]
  \centering
  \includegraphics[width=0.55\textwidth]{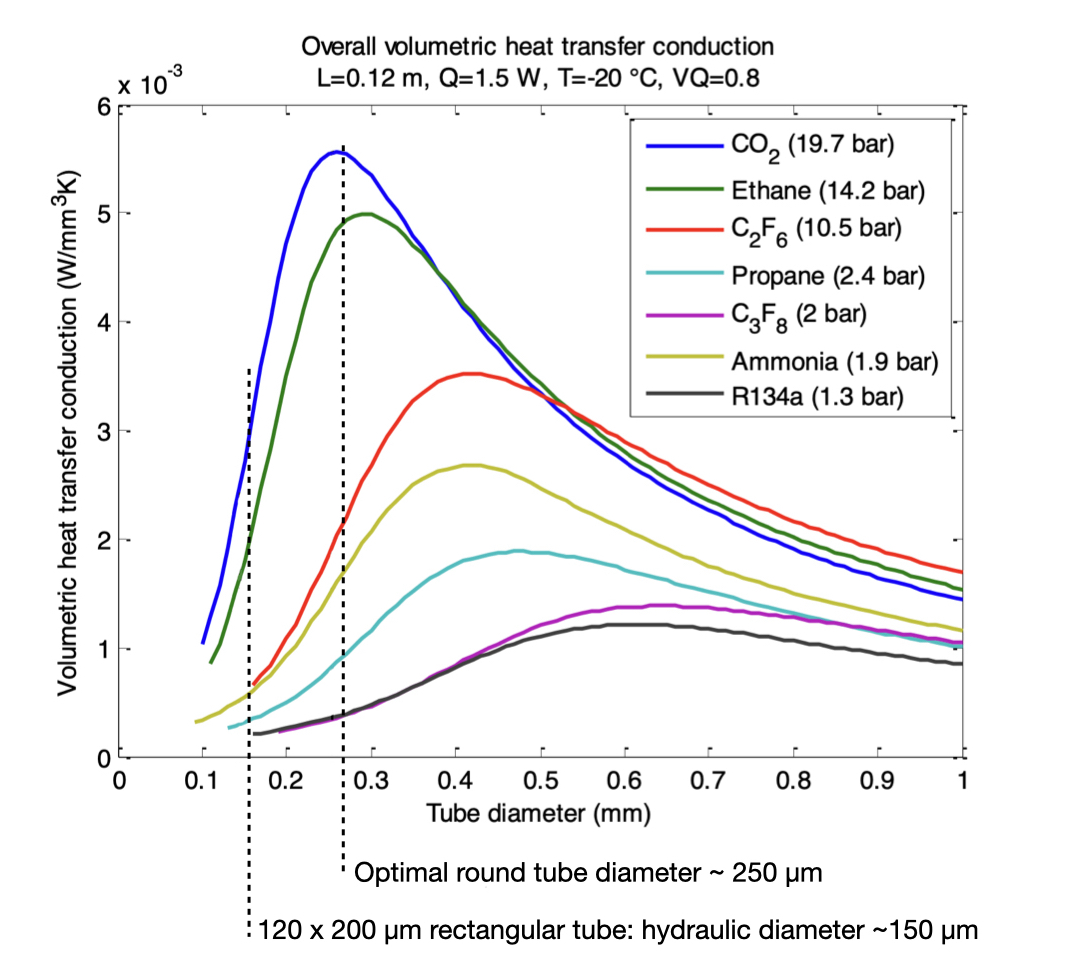}
  \caption{Result of CoBra simulation showing the heat transfer coefficient for various channel diameters and different coolants.  The optimum channel diameter is at $200 \mum$.  The final design choice for the LHCb VELO, driven by manufacturing and pressure resistance considerations corresponds to a tube hydraulic diameter of approximately $150 \mum$.}
  \label{cobra_simulation}
\end{figure}

The CoBra simulation program was used to analyse the expected coolant flow in the microchannels.  The results of this analysis were used to derive guidelines for the channel lengths and dimensions in the first prototypes.  CoBra has been developed to predict the complex behaviour of particle detector cooling circuits.  It is based on flow pattern models~\cite{thome1,thome2} which describe flow conditions and the related heat transfer and pressure drops, and include predictions of dry-out conditions.  A useful figure of merit is the volumetric heat transfer conduction, $\rm{\frac{Q}{V.dT}}$, where Q is the absorbed heat, V is the volume of the tube, and dT is the difference in temperature between the tube and the liquid.  The simulation was performed with model parameters of a tube length of 120 mm, absorbed heat of 1.5 \W, a temperature at the inlet of -20\degc and a vapour quality of the outlet of 0.8.  Coupling between channels was neglected and the tubes were assumed to be circular in cross-section.  The result of the simulation is shown in  \cref{cobra_simulation}.  As the tubes get smaller and the relative contribution of the wall area increases, the heat transfer coefficient improves, up to the point where the coolant viscosity limits the flow.  It is instructive to compare the performance of \cotwo with other coolant candidates.   It can be seen that \cotwo is well matched to small channels, and has an excellent performance due to its high latent heat and low viscosity.

\FloatBarrier

%% file: 05_microchannel_assembly.tex
\section{Microchannel assembly}

\label{sec:assembly}

\subsection{The assembly challenge}

The \cotwo is delivered to the microchannels via a connector with an integrated manifold and inlet and outlet tubes.  The connector must be attached to the silicon with a robust and leaktight seal, which is achieved via a soldering process.  The cooling connector also serves as the mechanical interface with the module support, in order to avoid introducing stresses on the microchannels by constraining them with a mechanical attachment.   Experience with the previously installed VELO detector has shown that where flux has been used in the cooling system construction there is a long term risk of corrosion, so the use of flux is not allowed.  The connector attachment must withstand the safety limit of~186 bar, must not block the channels, and must have no large internal voids which could compromise its high pressure resistance. 

The soldering process in which the cooler and its fluidic connector are attached proceeds in several steps. First, the cooler and the connector are pretinned in separate processes. The two components are then carefully aligned, and a reflow performed in order to form the connection.

\subsection{Connector design}

\begin{figure}[htbp]
    \centering
    \includegraphics[width=0.65\linewidth]{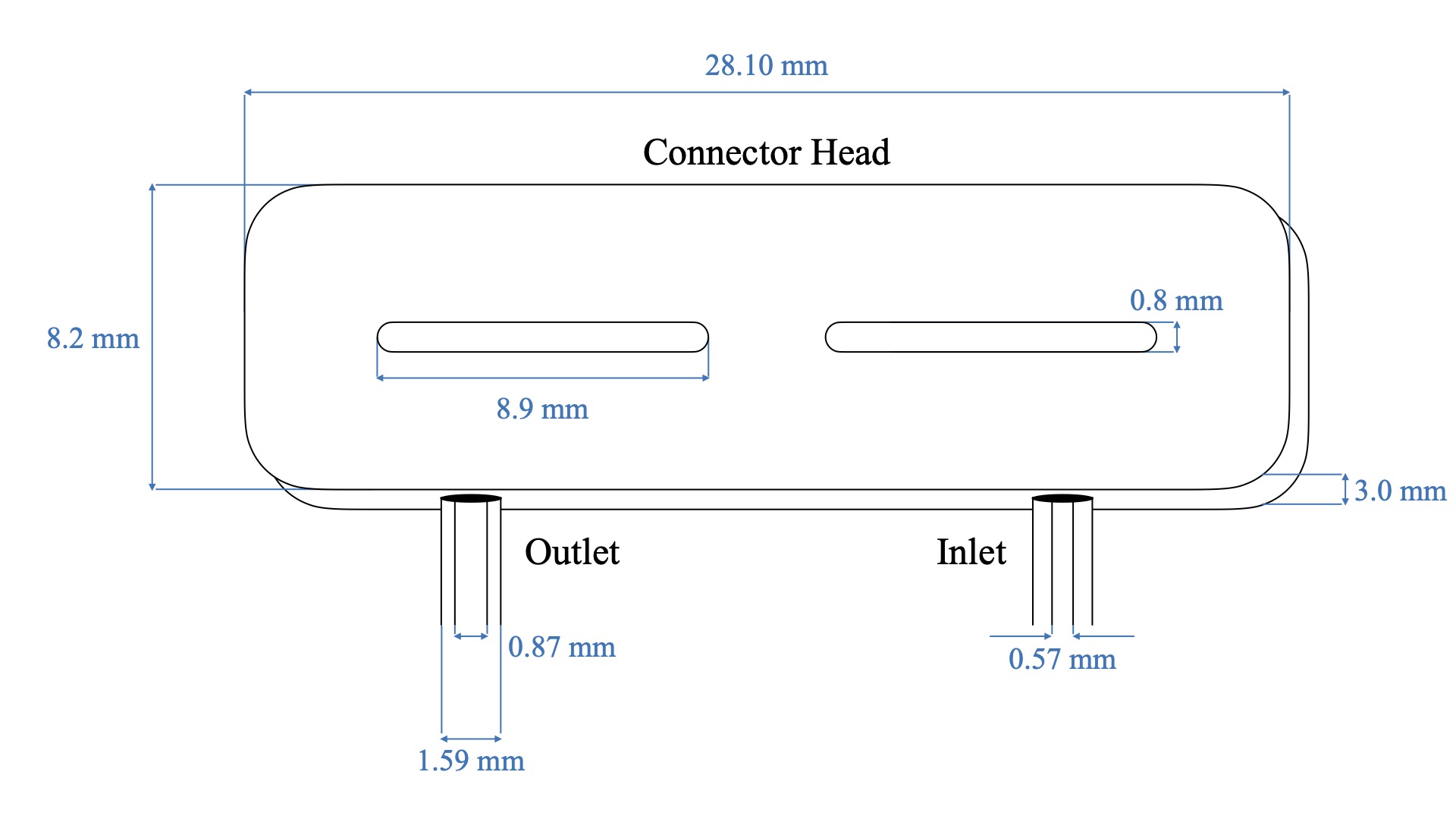}
    \caption{Illustration of a connector head with the tubes attached and relevant measurements indicated.} \label{fig:connector_design}
\end{figure}

The connector head is made from Invar 36 due to its coefficient of thermal expansion being close to that of silicon while the tubes and glands are made from stainless steel 316L. The connector head dimensions are $28.1 \times 8.2 \times 3.0$ mm with inlet and outlet slits of dimensions $8.9 \times 0.8 \times 2.1$ mm. The tubes have an unbent length of 318.6 mm and external diameter of $1.59 \mm$  with the inlet having an internal diameter of $0.57 \mm$ while the outlet tube has an internal diameter of $0.87 \mm$. The dimensions of the connector are shown in Figure \ref{fig:connector_design}. The connector head, tubes, and glands are vacuum brazed by CEA-CCFE. Rings of braze are placed into a custom-made jig designed to hold the connector head, tubes, and glands in the correct positions. The female nut is held in place around the tube while the tube and glands are brazed. During the brazing, the braze is above its melting point of 950\degc for 30 minutes.

The connectors are then tested to ensure they can withstand the safety limit of 186 bar, do not leak at the brazing points, and the inlet and outlet slits on the connector head do not exhibit large chamfers or burrs. First, the leak rate is measured with a helium leak test. The connector inlet and outlet slits are sealed by enclosing the connector head within two Invar blocks which are bolted together. Then the inlet tube is connected to a leak detector under vacuum and helium gas is sprayed around the brazing points. The pressure inside the inlet tube is measured over 10 minutes with a precision of $0.1\times 10^{-10}$ mbar. This procedure is repeated for the outlet tube. A pressure test is then performed, where the connector inlet and outlet slits are sealed by enclosing the connector head within two Invar blocks which are bolted together. Then the inlet tube is connected to a water pump, where the water pressure is brought to 200 bar for two minutes. The connector and tube are observed for signs of leaking or failure due to the pressure. This test is then repeated for the outlet tube and then the connector is flushed with pressurised air to remove residual water from the tubes and connector head. Lastly, a visual inspection with a high magnification microscope is performed to search for defects on the connector surface, the inlet and outlet, and the tubes. In particular, the connector is inspected to ensure chamfers at the surface edges, including the inlet and outlet, do not exceed a length of $100 \mum$. Additionally, any burrs found at the inlet and outlet are required to be $<50 \mum$ before passing inspection.


\subsection{Connector and cooler preparation}

Before being pre-tinned, the connector must be metallised. Experience showed that in order to ensure quality metallisation layers, a highly smooth and clean connector surface is necessary. Precise procedures were developed in order to obtain consistent results. First, the surface of the connector is polished in order to remove any ridges or depressions which might lead to inconsistencies or imperfections in the metallisation layer. The polishing is performed with increasingly fine sand paper. Visual inspection under a microscope is performed throughout the process. The connector must then be thoroughly cleaned. A sequence of ultrasound baths and rinsing with acetone, ethanol, and demineralised water is performed in a clean room environment. The connectors are then kept in a box filled with nitrogen until ready for metalisation.

\begin{figure}[htbp]
   \centering
  \includegraphics[width=0.99\linewidth]{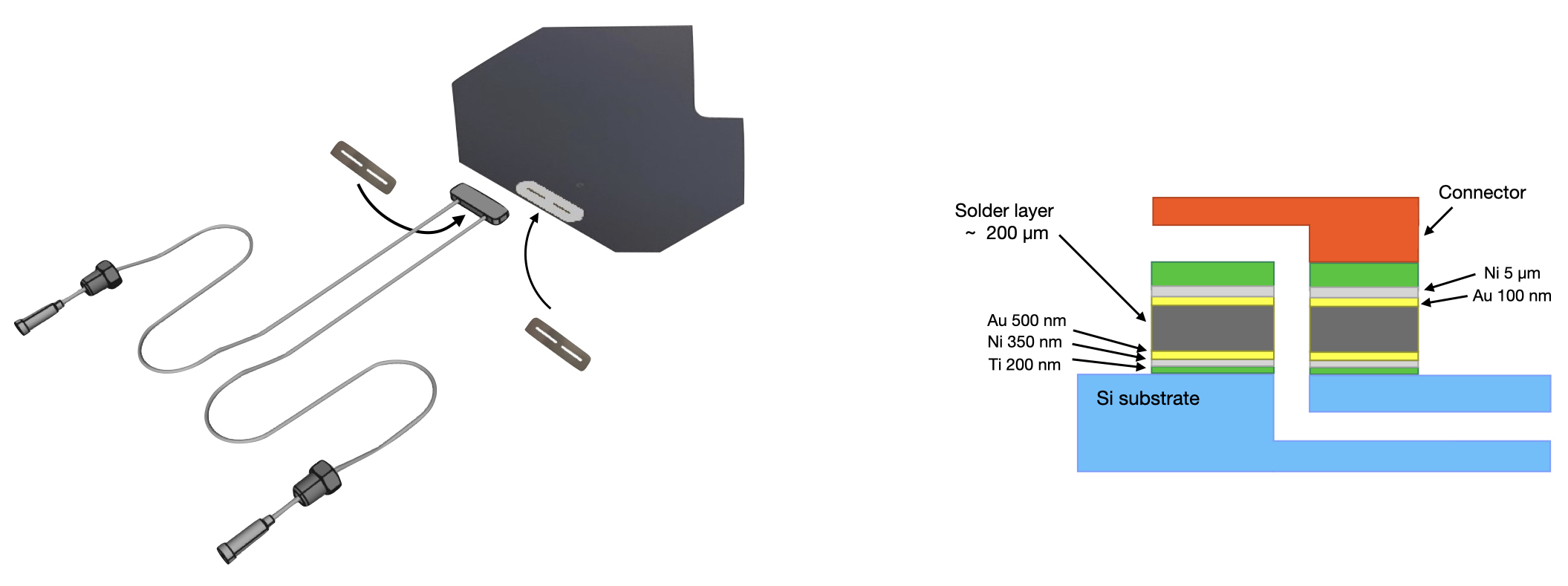}
   \caption{Illustration of assembly steps, with cross section of connector/cooler attachment} \label{fig:soldering_stack}
\end{figure}

The metalisation process consists of depositing a nickel layer of 5\mum followed by a gold layer of 200 nm. The nickel reacts and mixes with the solder to create the intermetallic compounds during the reflow, while the gold is used to prevent the oxidation of the samples and dissolves in the solder. Following this process, the metal layers are removed from inside the inlet and outlet regions and the connector is once again cleaned in an ultrasound bath and rinsed with ethanol. As described in \cref{sec:implementation}, the coolers have been previously metallised by LETI with a matching metallisation layer around the microchannel inlet/outlet region. Before pre-tinning, both the connector and the cooler undergo plasma cleaning in order to remove any possible organic contaminants. At this point, the components are ready to be pre-tinned. The metal stack is illustrated schematically in \cref{fig:soldering_stack}.

A $100 \mum$ thick Sn36Pb62Ag2 foil is placed on the metallisation layer, and the component left in a vacuum tank in order to outgas. After a period of at least two hours (twelve hours) for the cooler (connector), the tank is filled and flushed with nitrogen gas. Formic acid vapor is introduced in the nitrogen environment at a concentration of approximately 3\%. The sample is then heated to 150\degc and is left at this temperature for five minutes, in order to let the formic acid react with oxides on the foil and metallisation layer to create a salt or formate. These salts will decompose into $\cotwo$, water, and hydrogen when the sample is further heated above approximately 200\degc ~\cite{formicacid1,formicacid2,formicacid3,formicacid4}. The heating of the sample is initially performed slowly in order to maintain approximately equal temperature over the surface and a uniform melting of the foil. At around 183\degc, the foil is fully melted and the sample is heated rapidly to a temperature of 235\degc. The tank is put under vacuum in order to allow trapped gas bubbles to escape and then cooled down to room temperature.

\begin{figure}[htbp]
   \centering
  \includegraphics[width=0.99\linewidth]{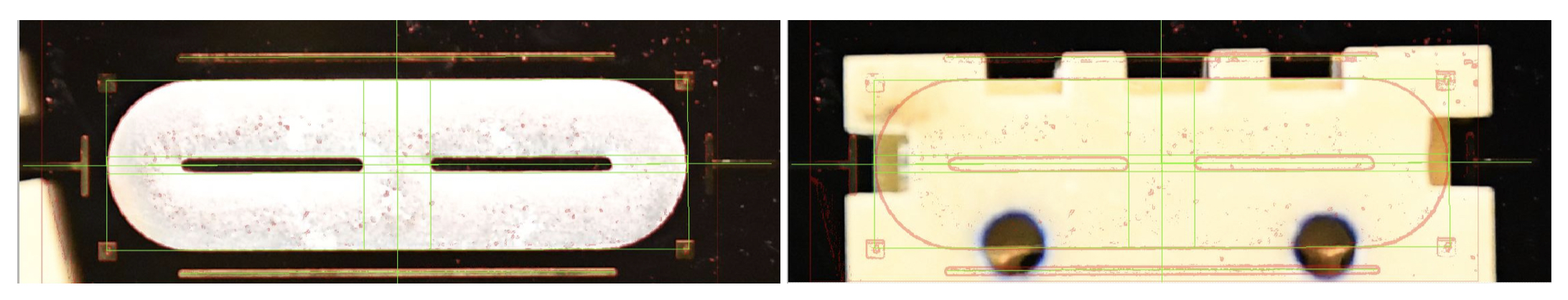}
   \caption{Images from the translational alignment of the connector with the cooler. Image processing tools are used to determine the position of the attachment region, and the connector is then aligned using these measurements. } \label{fig:mc_alignment}
\end{figure}

The pre-tinned coolers and connectors are visually examined to ensure no solder or other debris entered the slit regions. The components are mounted in the vacuum tank for the soldering, with the connector supported above the cooler by a stage which allows for micro adjustments of the rotation and translation with respect to the cooler. The connector is then carefully aligned over the matching pre-tinned area of the cooler using digital microscopes and cameras. Firstly, the rotational degrees of freedom are adjusted until the surface of the connector is completely parallel to the cooler. This is done with two cameras, one positioned in front and the other to the side. With the aid of image processing tools, a precision of $10-20\mum$ in the alignment of the edges is typically achieved. An overhead camera with image processing software is then used to perform the necessary translational adjustments. The process is demonstrated in \cref{fig:mc_alignment}. 

Once aligned, the connector is lowered to make contact with the solder layer of the cooler, which lies on an 8 inch silicon wafer used as a lightweight planar mechanical support capable of rapid heating in vacuum.  This wafer is itself supported by ball bearings to allow for self-correction of any remaining small misalignment. The oven is heated in vacuum at a rate of 1\degc/s to a final temperature of 235\degc, and the solder layers melt to form the connection. Before cooling down, the tank is pressurised with atmospheric air in order to compress any voids in the solder. The sample is then cooled down to 100\degc, at which point vacuum is reintroduced, and the same procedure repeated in order to reduce the chance of any lingering voids. 

\begin{figure}[htbp]
\centering
  \includegraphics[width=0.7\linewidth]{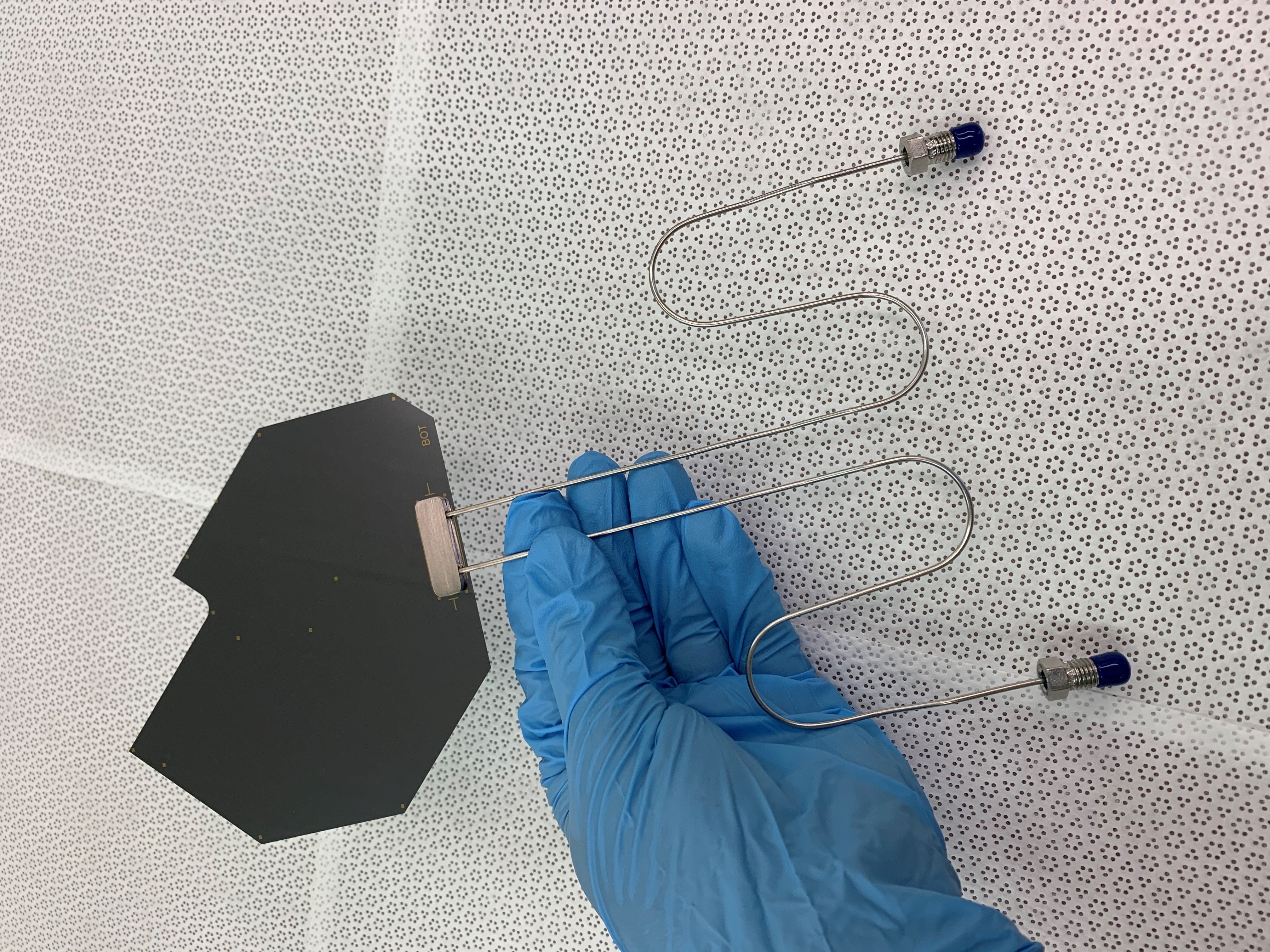}
   \caption{Microchannel assembly, consisting of a microchannel cooler soldered to a fluidic connector, ready to be equipped with VELO module components.}
   \label{fig:MC_final_assembly}
\end{figure}

The final step of the assembly consists of gluing a rectangular piece of silicon onto the cooler on the side opposite the connector. This additional thickness serves as reinforcement against the pressure of the CO$_2$ flowing from the connector into the cooler.   A photograph of a finished microchannel assembly is shown in \cref{fig:MC_final_assembly}.

\subsection{Alternative cooling solutions}
\label{sec:alternativecoolingsolutions}

The microchannel technology development, both for the cooler production and the assembly process, was a tremendous challenge for the VELO team.  In case the R$\&$D would not have been successful, or the assembly yield too low, backup options were also developed (see~\cite{belleiiupgrade,freekdetectorseminar}).  One option consisted of machining precision grooves in a ‘‘Shapal" (aluminium nitride) plate into which stainless steel pipes were embedded, after robot deposition of heat conductive glue. The Shapal plate provided mechanical support and excellent thermal conductivity.  Orifices for the expansion step were provided in an inlet block grouping all the channels.  A second solution consisted of manufacturing 3D printed titanium substrates.  Such an approach is in principle cheap and provides a flexible layout and fast turnaround for prototyping. It benefits from vast experience in industry and the material can be welded or brazed to the cooling pipes.  Inlet restrictions can be easily incorporated into the design.  At the time of the VELO development feature sizes of $100 \mum$ were readily available in a suitable material, and this alternative reached a mature stage. Other proposals included integrating the cooling pipe into a ‘‘pocofoam" substrate, force fitting a cooling tube into TPG sheets, or even the use of a diamond substrate, however after initial prototyping these ideas were not taken further.  After internal review the silicon microchannel solution was chosen for the VELO upgrade, however additive manufacturing techniques are currently under active development for future applications~\cite{Aglieri:2764386}.

%% file: 06_microchannel_qa.tex
\section{Microchannel quality assurance} 
\label{sec:qa}

The qualification procedure must ensure that the microchannel assemblies can be safely installed inside the secondary vacuum of the detector, and that there is no obstruction which could cause a reduction of CO$_2$ flow. Therefore modules are extensively checked for leak tightness and overall robustness. Long-term effects in the solder joint such as creep and fatigue are tested with a subset of prototypes. 

\subsection{Microchannel assembly quality assurance}

\begin{figure}[!htb]
\centering
  \includegraphics[width=0.45\linewidth]{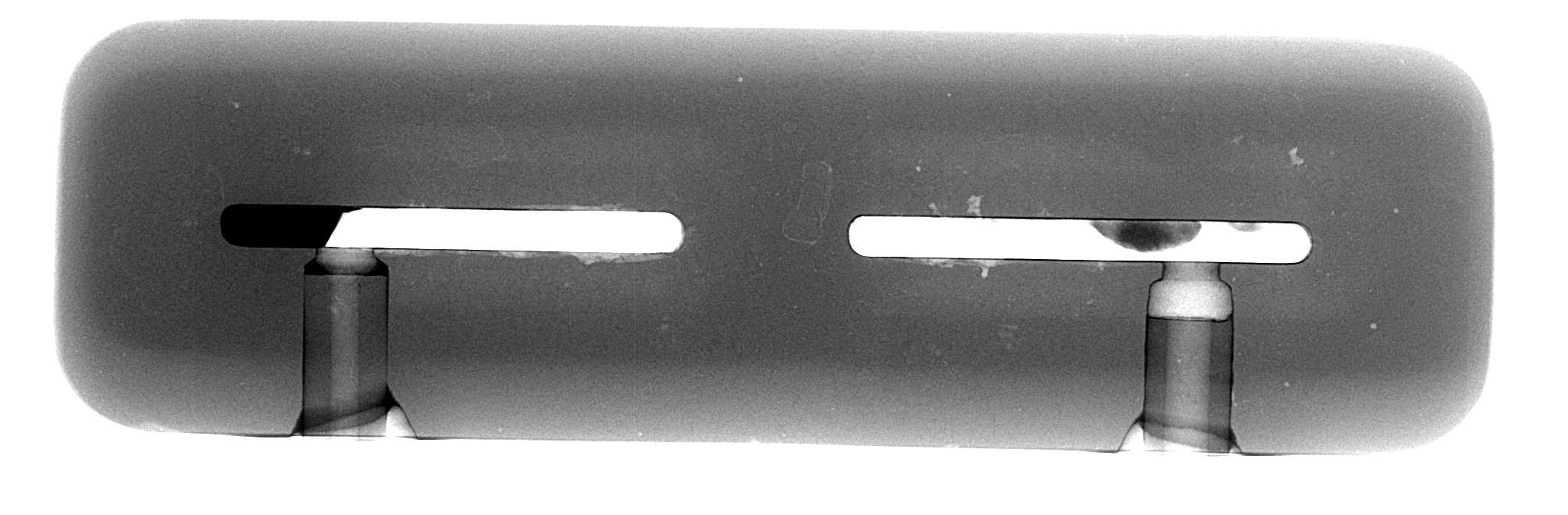} \hfill
  \includegraphics[width=0.45\linewidth]{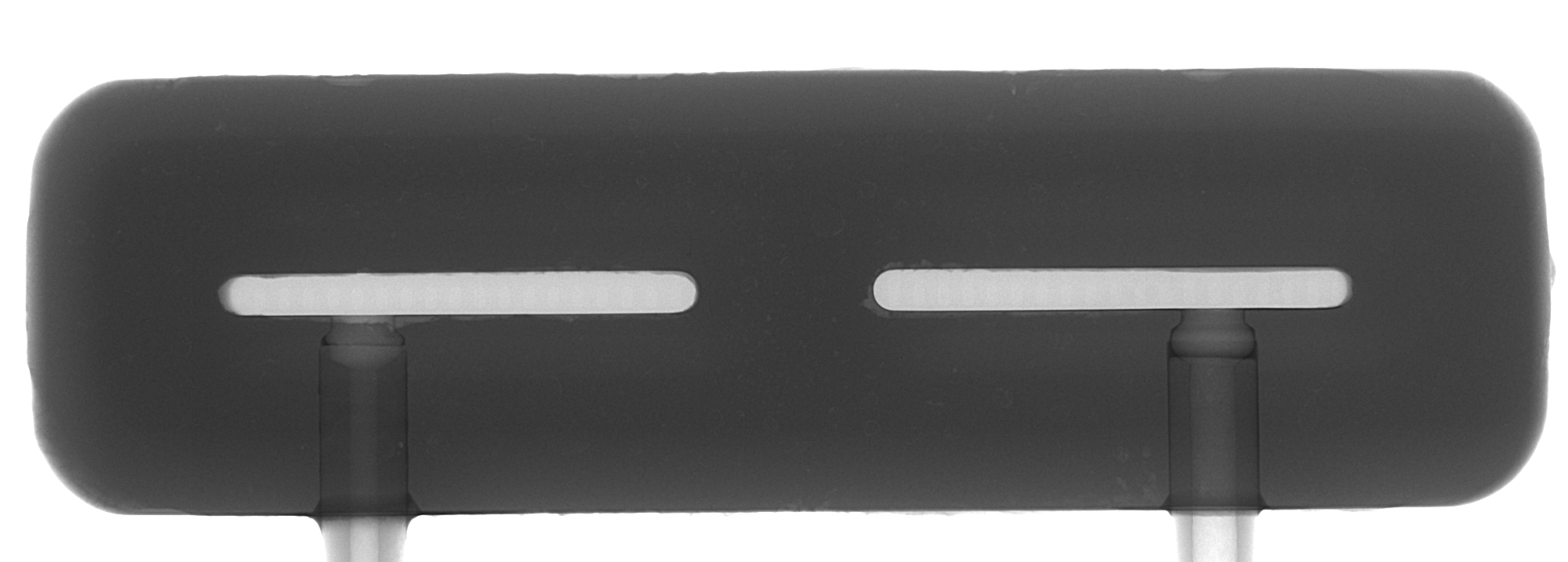}
   \caption{X-ray of solder joints attaching the fluidic connector to the microchannel cooler. The left image demonstrates an unsuccessful soldering in which solder has crept into the slit regions. Voids connect to the slits are also visible. The right image demonstrates a successful soldering, with no voids or solder excess. }\label{fig:mc_xray}.
\end{figure}

The quality assurance begins with an X-ray of the solder joint. This test allows for verifying that solder did not enter the inlet or outlet region, and that there are no large voids in the joint. The absence of such features rules out the possibility of the presence of obstructions which might affect the CO$_2$ flow and potential weak points which might compromise the integrity of the joint. Examples of the X-rays can be seen in \cref{fig:mc_xray}, both for the case of a failed and successful soldering. If solder access into the slit region is observed, a 3D X-ray tomography is performed in order to determine if the channels are blocked. The tomography obtained typical resolutions of $\sim 15 \mum$.

Before subjecting the sample to the rigorous robustness tests, it is first checked for leak tightness. A leak tester is attached to the fluidic connector and creates a vacuum inside the sample. Helium is sprayed all around the cooler while a gas spectrometer is used to detect any helium entering the vacuum. The procedure excludes leaks down to approximately $10^{-9}$~mbar~l/s. To check for robustness, the samples are subjected to high pressure tests. The system is initially pressurised to 130 bar, which corresponds to the burst disk release pressure installed in the final system. High purity nitrogen is used for the pressurisation in order to ensure internal cleanliness. After 45 minutes, the pressure is increased to a final value of 186 bar. This last step is done to include an additional safety buffer in the qualification, and lasts for 15 minutes. A final high-pressure leak test is performed. The microchannel assembly is installed inside a vacuum chamber and is pressurised to 60 bar with helium while a gas spectrometer is used to detect any leakage of helium into the vacuum chamber. The typical background level is around $10^{-7}$~mbar~l/s and no significant leak rate increase is observed when the microchannel assembly is pressurised.

\begin{figure}[htbp]
\centering
  \includegraphics[width=0.8\linewidth]{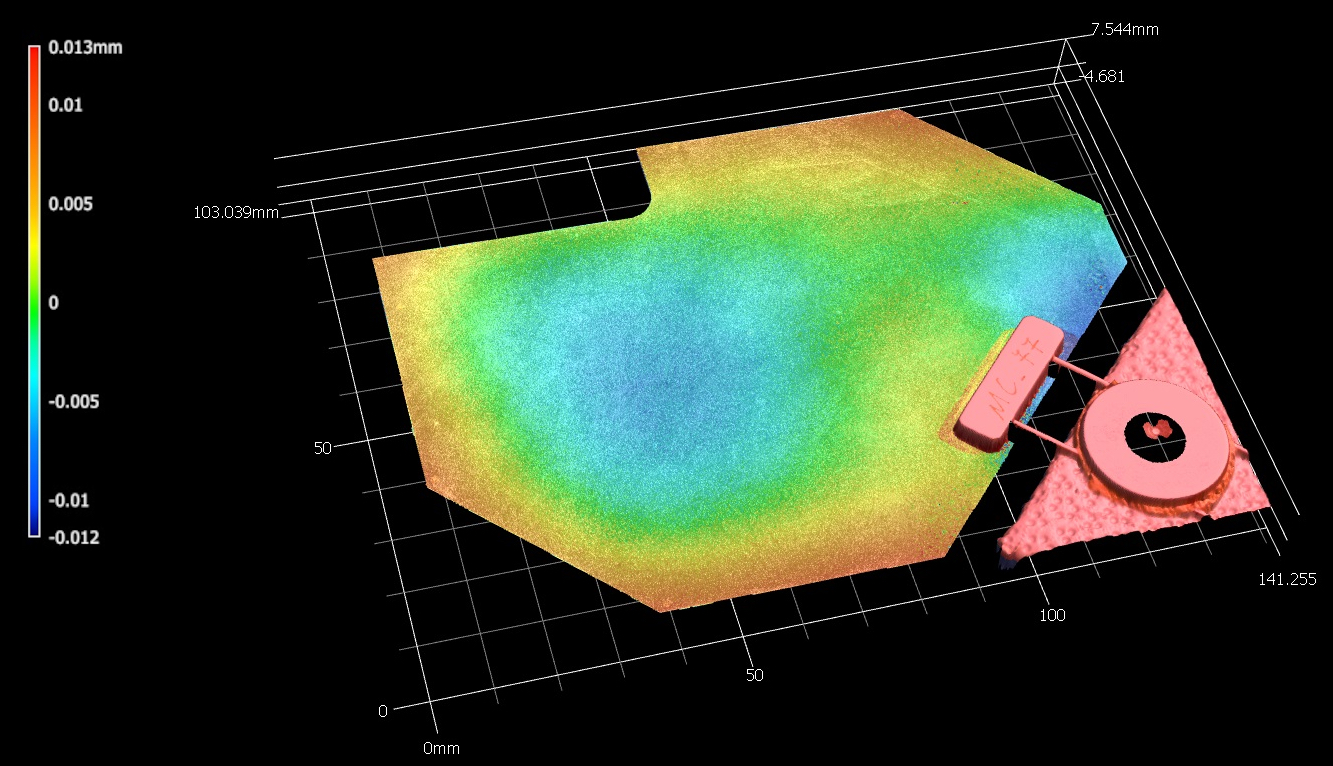}
   \caption{Microchannel surface survey after the attachment of the fluidic connector. The blue(red) regions indicate the lowest(highest) regions. The overall variation in this case is around 25\mum over the surface of the cooler.}
   \label{fig:keyence}
\end{figure}

For the subsequent module gluing and assembly steps, the silicon surfaces must also meet flatness requirements. The planarity is measured using light reflectometry devices capable of reaching an accuracy of $\pm3$\mum. As the silicon has a reflective surface, a fine layer of talc is applied in order to be able to resolve the surface. The tests are performed before and after the connector attachment, in order to confirm stress from the soldering procedure did not introduce deformations. An example of these measurements can be seen in \cref{fig:keyence}. The maximum deviations from planarity never exceeded $100 \mum$. 

The silicon surface must also be clean and free of any surface residue. The cooler is rinsed with ethanol and a final visual inspection is performed. The shape of the connector capillaries must also conform to predetermined dimensions, and is checked using a custom 3D jig.

\subsection{Creep and fatigue}

Over time the solder joint may be subject to failure due to sustained strain over a long period leading to plastic deformation.  Creep effects can typically be neglected if the temperature is less than $T_{\rm critical}$, defined as 50\% of the absolute melting temperature.  For the $\rm SnPb$ solder used this corresponds to $T_{\rm critical}$ = -46\degc. The VELO will operate for nearly a decade, and will be maintained at -30\degc for the majority of its lifetime. Therefore, creep effects are expected to be small.  Using the model in ref.~\cite{10.1115/1.1462624} the strain rate was simulated for the solder joint, taking into account two scenarios, one with the stress from the \cotwo pressure spread uniformly over the joint, and one with the stress concentrated around the slit.  It was found that at -30 \degc the time to fail for the worst case was approximately $10^6$ days, dropping to approximately $2 \times 10^4$ days at room temperature. During the test, the sample pressure was kept at 60~bar to mimic the \cotwo pressure at 20\degc while the temperature was kept between 90-120\degc to accelerate the creep process. In total, 15 samples were tested during a testing campaign of approximately three months (2477 hours). Two failures were observed with samples which were prepared with flux and none of the samples prepared with the fluxless procedure failed.

The potential effects of fatigue on the joint were also considered, corresponding to a weakening of the joint due to repeatedly applied loads.  During the VELO lifetime a few hundred temperature cycles are expected, corresponding to powering during the LHC fills.  In order to check the effects of fatigue a dedicated test stand was set up.  Microchannel samples were connected with a high pressure connection and mounted on a heating/cooling plate equipped with six $40 \W$ Peltier coolers.  The temperature was cycled between -40\degc and +60\degc and at each point in the cycle it was possible to cycle the pressure between atmospheric pressure and $186 \bar$.  The fatigue effect was evaluated with nine samples. Overall, approximately, 6800 temperature cycles and 40000 pressure cycles were performed over several months. The pressure cycles were done only when the sample was at high temperature to avoid accumulating ice inside the channels while the sample is below 0\degc. One failure was observed on a sample with flux. The post-mortem microscopic photos showed that there were regions which were not properly attached due to a thin layer of flux between the connector and silicon interface.

\begin{figure}
\centering
\includegraphics[width=0.99\linewidth]{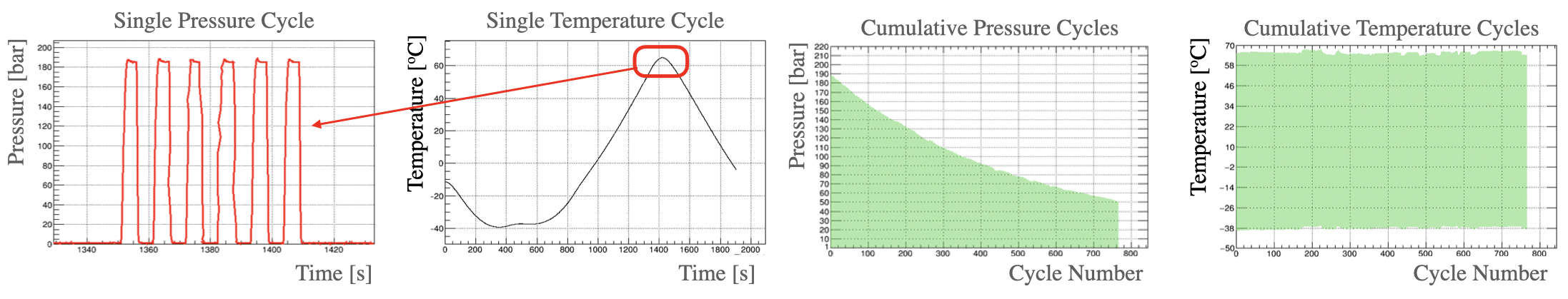}
\caption{The two plots on the left show recorded data during one temperature cycle, lasting around 30~min. At the warmest temperature (around 60\degc) a pressure cycle is performed.   The two plots on the right show data from multiple cycles. The min-max difference is shown for the pressure and temperature respectively from left to right. It can be seen that the pressure provided to the sample decreases over time since the reservoir becomes depleted with every pressure cycle.}\label{fig:fatigue_test_stand}.
\end{figure}


\subsection{COVID procedures}

The microchannel cooler is the backbone of the detector modules, and a constant supply of the microchannel assemblies was critical for the timeliness of the detector construction. The COVID-19 pandemic began in the middle of the construction period, and presented several organisational challenges in assuring a reliable supply of coolers. 

One major aspect of this was the interruption of travel, which required adaptation in transportation of materials to avoid supply chain disruptions. Travel between the different sites was originally frequent enough that components could simply be carefully packed in personal luggage. With the advent of the pandemic, it became necessary to rely on more conventional shipping methods. Particularly in the case of the assembled microchannel coolers, this required development of secure shipping methods that will be described in detail in the next section. 

Compliance with social distancing guidelines required careful planning in the lab spaces. Restrictions elsewhere meant limited availability of several services such as connector metallisation, x-ray tomography, and clean room access. These complications required increased parallelisation of procedures. Originally, many parts of the process were done one-by-one, such that if a problem arose, for example with the metallisation, not many components would be affected. With the above complications, such a system was no longer possible and so more components needed to be processed simultaneously. In order to not risk losing batches of the limited supplies to unidentified problems, stricter quality control was enforced to ensure that problems were identified at an early stage.

Despite these obstacles, the above adjustments allowed for a constant and reliable production of the coolers. The production had a high yield, and did not result in any significant delays with respect to other elements of the detector.

\subsection{Packaging and transport}

\begin{figure}[htb]
\centering
  \includegraphics[width=0.7\linewidth]{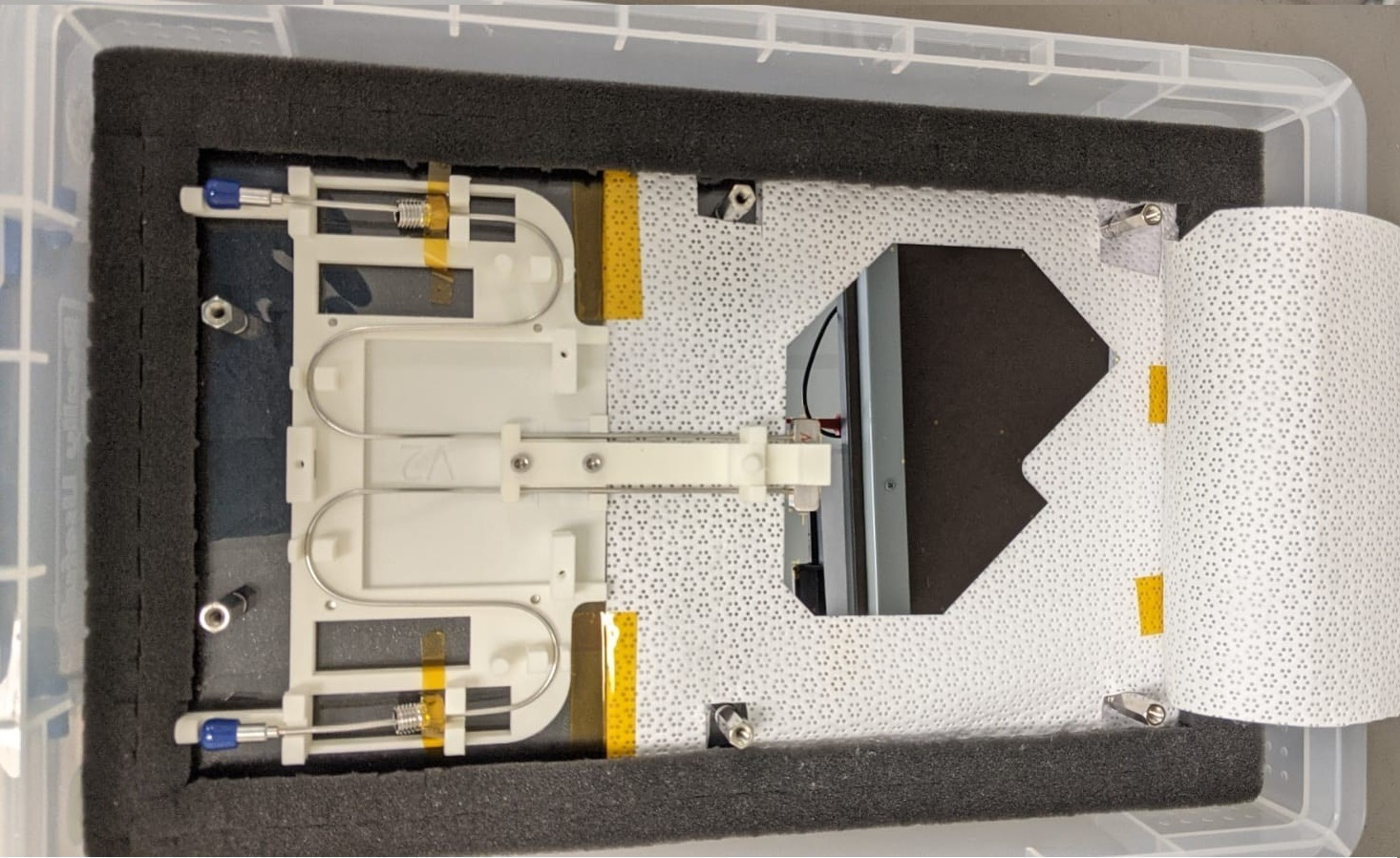}
   \caption{Microchannel assembly mounted in the transport jig.}
   \label{fig:MC_packing}
\end{figure}

After the qualification and final visual inspection the coolers were packed into small transport boxes as illustrated in \cref{fig:MC_packing}. The tubes were held in place by a 3D printed jig and the fluidic connector was gently pushed down against a base plexiglass plate which is covered by a soft clean room grade paper. Each small transport box could carry up to two microchannel assemblies and were typically transported as hand luggage either by plane or car. Later on, those boxes were integrated into a  larger transport box fixed on a pallet to allow the usage of standard air freight. The large transport box was quite oversized for such small structures and had at least 9~cm of polyethylene foam around each of two small transport boxes. No microchannel assemblies were damaged during the transportation to the module assembly sites. 

\subsection{Assembly yield}

The fluxless soldering procedure was refined and tested extensively using lower grade coolers in order to perfect the procedure.  As the modules will be mounted in the secondary vacuum of the LHC, only those coolers which perfectly passed all QA procedures were accepted for installation. Firstly, the procedure was perfected with the manufacture of approximately 30 prototype coolers, which had the additional benefit of providing a plentiful supply of coolers for module assembly prototyping. Next, the production phase was started.   Over a period of 27 months, including the COVID lockdown period, 81 installation quality microchannel assemblies were produced with a yield of 87\%.   Of the assemblies which were deemed to be not of the highest quality for installation, analysis shows that eight could be recovered, and these are held in reserve.

%% file: 07_cooling_performance.tex
\section{Cooling performance} 
\label{sec:cooling}

\subsection{Characterisation and operational stability testing}

\begin{figure}[ht]
 \centering
     \includegraphics[width=0.99\textwidth]{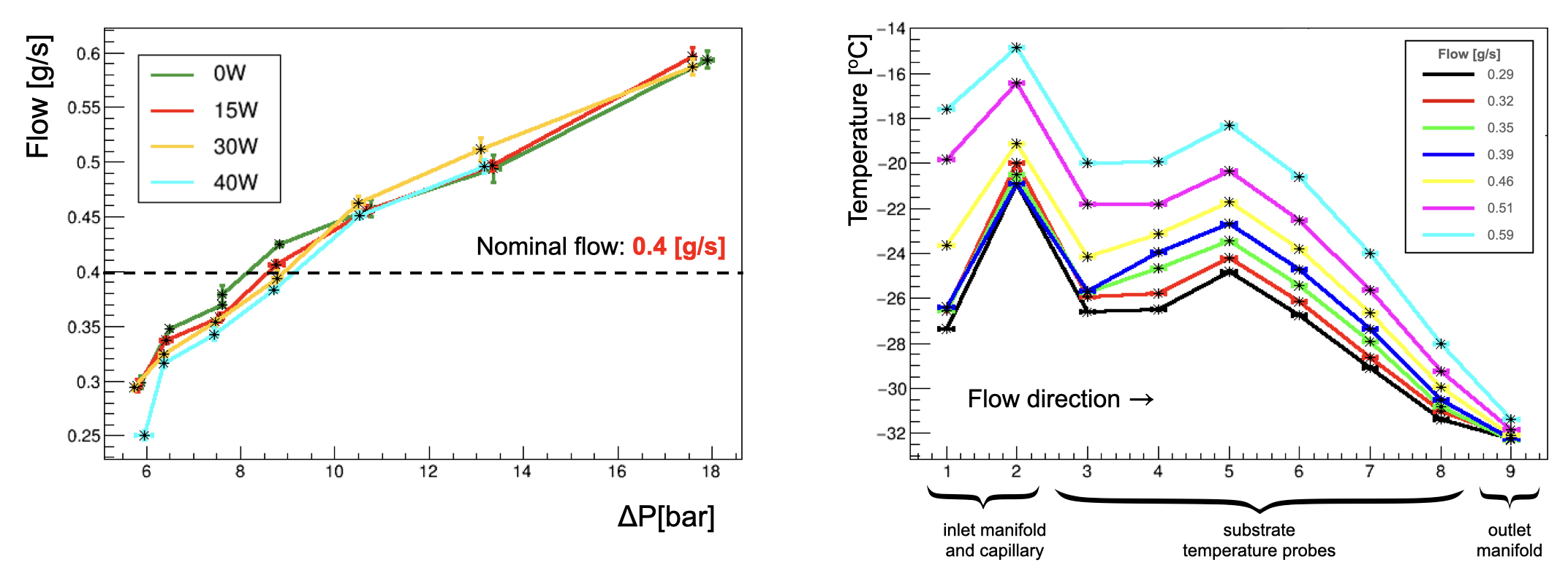}
    \caption[]{Left plot: Measured flow as a function of pressure drop over the cooler.  The curves are similar for a range of power dissipation, indicating good operational stability.  Right plot: Temperature evolution from the inlet to the outlet for a range of power dissipations where the probes from left to right indicate increasing distance along the channels.  Overall there is a drop in temperature due to the pressure drop through the channels.}
  \label{fig:fluidic_characterisation}
\end{figure}

Two microchannel assemblies were used to measure the full fluidic characterisics, thermal performance and stability under joint running with a \cotwo cooling plant called CORA (\cotwo Research Apparatus)~\cite{CORA}. They were each equipped with a full set of 12 silicon heaters to simulate the ASIC load in the final VELO module and an additional silicon heater to simulate the sensor load in the region where the sensor overhangs the cooler and suffers the highest radiation. The temperature was monitored with nine temperature probes on each cooler, tracking the temperature evolution from the inlet through to the outlet manifold.  In order to be as close as possible to the final configuration, the \cotwo was supplied through a capillary with a length similar to that in the VELO tank, equipped with a $150 \mum$ gasket to control the flow as is installed for the final assembled VELO. Results from the cooling test are shown in \cref{fig:fluidic_characterisation}.  On the left, the flow as a function of the pressure drop over the cooler is shown.  The accumulator \cotwo temperature set point was -35\degc and it was checked that the boiling was correctly triggered in the microchannels.  It can be seen that there is a very small spread between the curves with the module set to a range of power dissipations between zero and the maximum.  This gives a good indication that the system will avoid instabilities in the case of parallel operation of many modules which may have different power dissipation.  At the nominal flow of 0.4 g/s the pressure drop is approximately 9 bar. The plot on the right gives an indication of the temperature evolution along the length of the cooling channels.  The highest temperature is seen at the inlet.  After passing the point where the boiling is triggered, the temperature then drops along the length of the channel.  This is due to a small drop in pressure, due to the additional resistance created by the boiling liquid.

\begin{figure}[ht]
 \centering
     \includegraphics[width=0.95\textwidth]{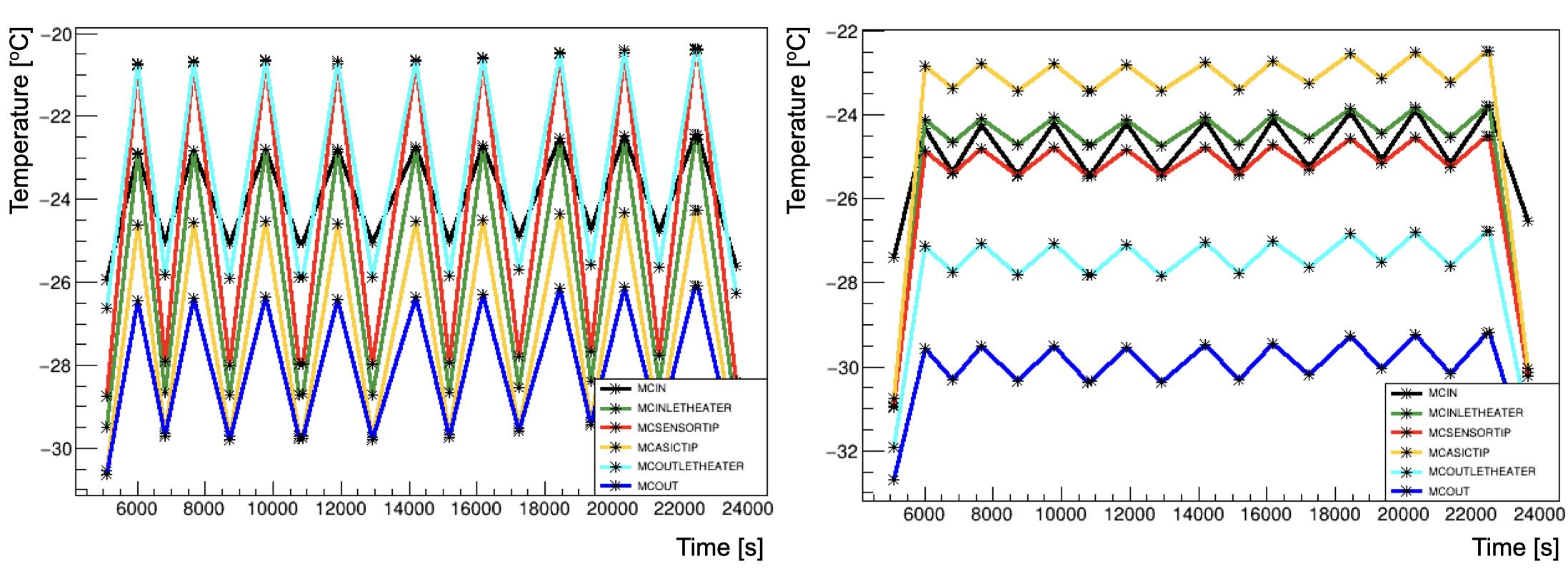}
    \caption[]{While keeping the constant nominal power dissipation of 30W on microchannel B (right plot), the power on the microchannel assembly A (left plot) was switched on and off multiple times to test the stability of the boiling. Both plots show the temperature on the coolers in different locations as a function of time.}
  \label{fig:stbility_test}
\end{figure}

A stability test was also performed to check the interplay that might be expected between two VELO modules in the case of changes in fluidic resistance due to changes in power.  Therefore, the initial tests were performed with the two equipped microchannel assemblies in parallel. While keeping the power constant in one assembly, the second one was cycled from unpowered to the nominal expected power several times as  can be observed in \cref{fig:stbility_test}. During this specific run, the power on the microchannel assembly B was kept at 30W while the power on microchannel assembly A was switched on and off (from 0W to 30W) multiple times. Overall, a small variation of temperature was observed on the microchannel assembly B (order of 1\degc or less) and, more importantly, the boiling was maintained throughout the power cycles.

\begin{figure}[ht]
 \centering
     \includegraphics[width=0.89\textwidth]{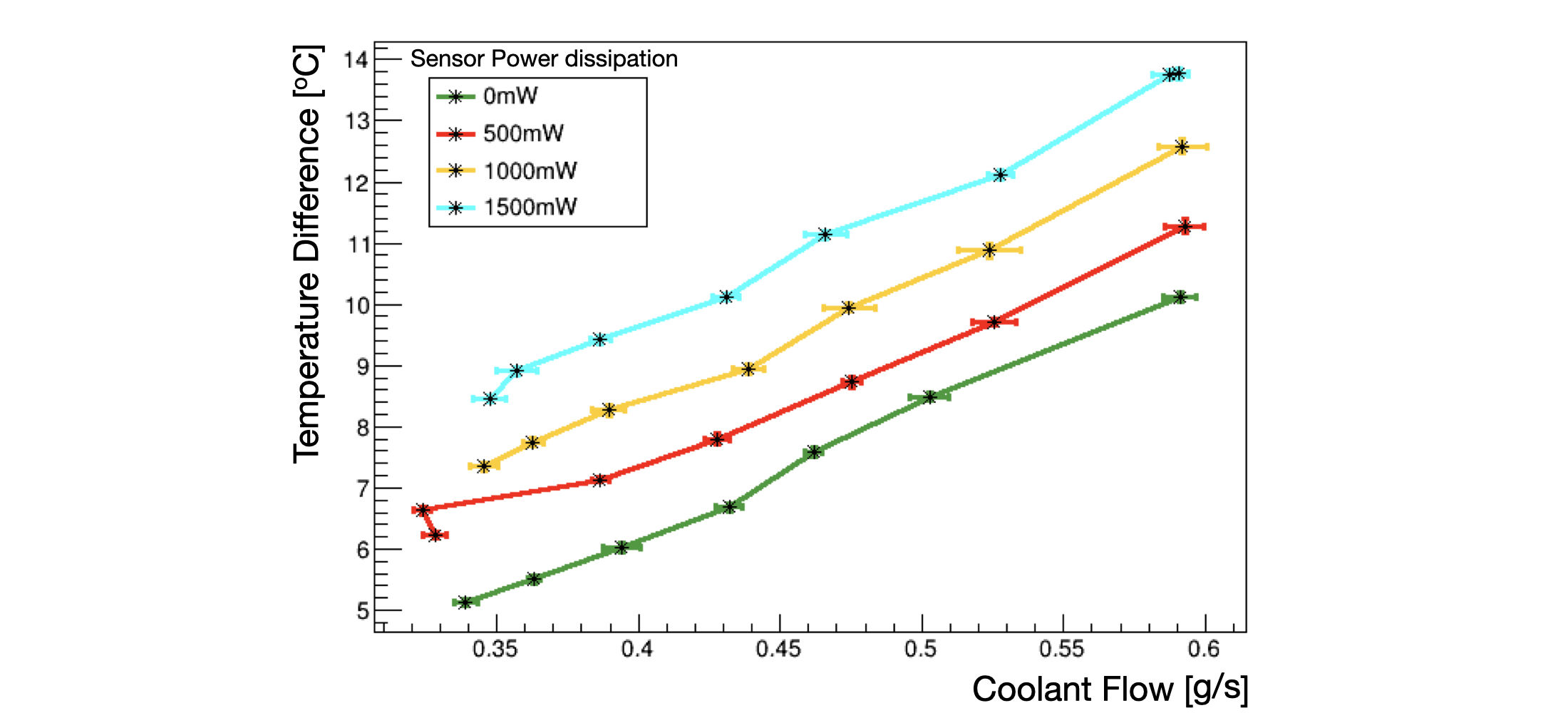}
    \caption[]{Temperature difference between the outlet manifold and the tip of the sensor - which extends beyond the microchannel cooler - as a function of coolant flow.  Four curves are shown, corresponding to the expected sensor power dissipation from the start through to the end of lifetime of operation at the LHC.}
  \label{fig:cooling_characterisation}
\end{figure}

Finally a measurement was performed to check that the temperature at the very tip of the module.  Here the silicon overhangs the microchannel and so is not directly cooled.  This temperature difference is expected to increase as the irradiated tip draws more current at the end of lifetime.  The measurement was performed as a function of coolant flow for four different power dissipations.  The results, displayed in \cref{fig:cooling_characterisation}, show that the temperature in the sensor can be kept comfortably below the target of -20\degc for the nominal power dissipation of $1 \W$ and coolant flow of 0.4 g/s.

\subsection{Construction of VELO modules using microchannel assemblies}

The microchannel assemblies finally undergo a series of construction steps in order to attach the active components and services to complete the VELO modules.
This involves a sequence of complex handling procedures.

In a first step the microchannel assemblies are glued, via the connector, onto a carbon fibre mid-plate, which allows the position of the cooler to be accurately defined with respect to the module base, to a level of $50 \mum$ in the glue thickness direction (beam direction) and a nominal $20 \mum$ in the transverse direction (module height).
For this step the edges of the cooler, butted against alignment pillars on the jig, are used.
Next, the hybrid pixel tiles and front end hybrids are glued to the cooler, using Stycast FT2850 with catalyst 23LV for optimum heat transfer for the tiles, and Loctite SI 5145 for the hybrids where more flexibility is required due to the CTE difference.
This is done in a double sided process, with the module supported in a turnplate, attached by the foot and with additional support for the cooler, sandwiched between independent vacuum jigs supporting the tiles.  The alignment marks on both sides of the cooler are useful to relate the position of the cooler and the elements glued on it with respect to the jig.
The electrical connections between the hybrids and tiles are formed with wirebonds, with support provided by custom-made mechanical supports below the coolers.  

Extensive metrology and visual qualification is performed after every assembly step, the final module is then electrically and thermally evaluated in vacuum.
The full suite of tests is repeated again after a thermal cycling consisting of ten cycles of cooling and heating from room temperature down to -30 \degc and back.
More details about the module assembly steps can be found in ref.~\cite{Svihra:2727215}.

\begin{figure}[htbp]
 \centering
     \includegraphics[width=.95\textwidth]{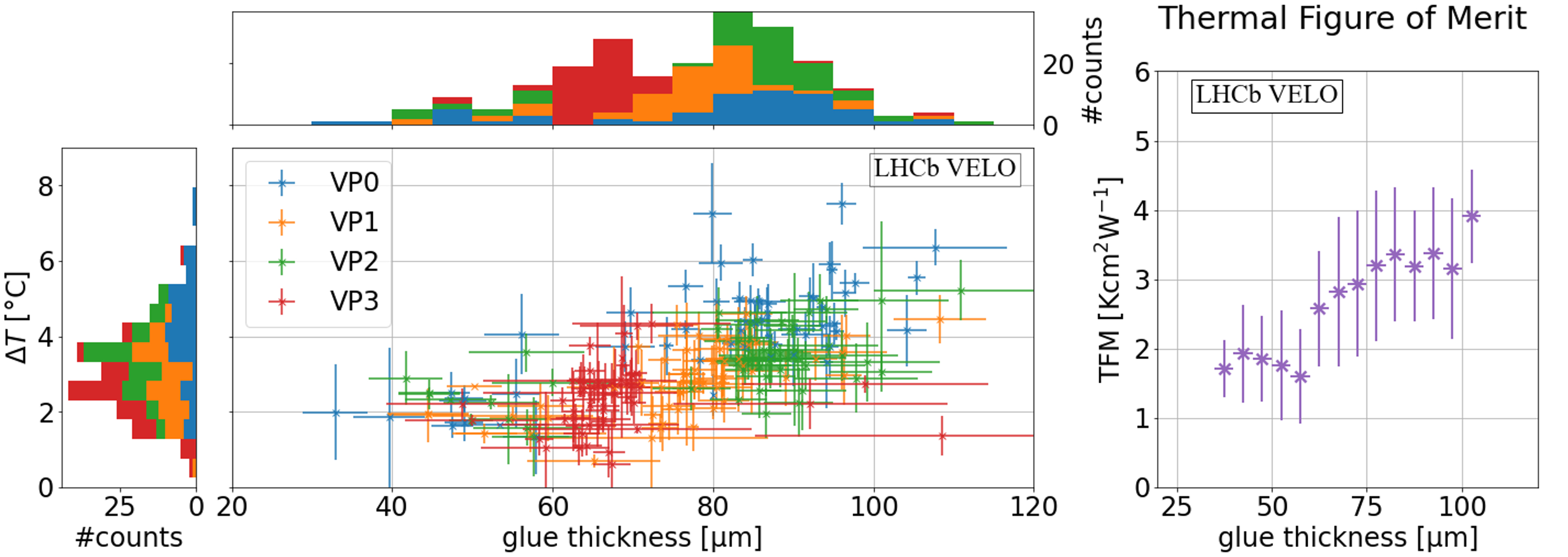}
    \caption[]{Left plot: change in temperature of the four tiles on the modules (VP0 to VP3), when going from zero to full power (26$\W$). The values are derived from the ASIC bandgap measurements after thermal cycling and each tile is assigned the average value of the three ASICs.  The x-axis gives the metrology measurement of the glue thickness for each tile.  Right plot: Thermal figure of Merit (TFM) for the VELO modules, derived from the temperature and power measurements of each ASIC and plotted as a function of glue thickness. The values are binned with $5 \mum$ thickness steps and a constraint on ASIC thickness variation of $<5 \mum$ is applied. All errors represent standard deviation of the measurements.}
  \label{fig:cooling_afterTC}
\end{figure}

Throughout the assembly of the first 52 modules the microchannel assemblies have been shown to be robust against these handling procedures.
This is best shown on bandgap\footnote{internally implemented temperature measurement in the integrated circuit} measurement from the ASICs (see \cref{fig:cooling_afterTC}), where, even after thermal cycling, all tiles at the end of life of the module are expected to observe a temperature difference less than 8 degrees from the coolant.  The difference in $\Delta \rm{T}$ between the tiles is dominated by the thickness of the glue layer under the ASICs, as obtained from the module metrology.  
It can be concluded that the thermal performance is excellent and is homogeneous across the module cohort. 

These data can also be used to measure the Thermal Figure of Merit (TFM) for the system, by monitoring the change in temperature on each ASIC, as measured by the internal band gap, as the module power is increased from zero to 26\W.  At maximum power, each tile consumes about 5.7 W, hence has a power density of 0.88~${\rm W/cm^2}$ (1.48~${\rm W/cm^2}$ is taken for the overhanging tiles due to their reduced contact area with the cooler). The data from 680 ASICs was used to calculate the TFM in bins of glue thickness. The glue thickness is derived for each tile from an average of more than 100 metrology measurements with a standard deviation of less than 5\mum for the ASICs used for this analysis.  The result is shown in \cref{fig:cooling_afterTC}. It can be seen that the TFM has a dependence on glue thickness and for the thinnest glue layers TFM values of between two and three are achieved, measuring on the final modules as constructed.  The LHCb tile attachment uses on a star shaped glue pattern to eliminate the possibility of encapsulated voids in the vacuum system, and is not designed for thicknesses much below 50\mum.  The glue thickness is an effective glue thickness as measured which may contain some empty areas.  A uniform glue coverage may improve the TFM still further.  This result showcases the high efficiency of the microchannel cooling design. For comparison the best measured TFM to date on a detector system installed at the LHC is $ 13\rm{~K~cm^2 W^{-1}}$ for the integrated steel pipe cooling system of the ATLAS IBL\cite{Capeans:1291633}, while standalone microchannel and integrated test chips have been shown to achieve TFM values of $3-4 \rm{~K~cm^2 W^{-1}}$\cite{Hellenschmidt:2689031,3dverticalintegrated}, consistent with the LHCb VELO module measurement.

%% file: 08_summary.tex
\section{Summary and outlook} 
\label{sec:summary}

The microchannel assembly production for the VELO Upgrade I pixel detector has been successfully completed.  The necessary $\rm{R\&D}$ to bring this emergent technology from concept to production stage has been described.   The microchannel coolers circulate bi-phase \cotwo, allowing the VELO detector to operate at very high beam intensity, in vacuum, with ultra light weight and radiation hard cooling.  The microchannel coolers also provide mechanical support for the sensors, front-end electronics, and hybrids.  The cooling is efficient enough to allow the sensors to overhang at the module tip, hence there is no material budget contribution from cooling for the first few $\mm$ of track measurements.  The microchannel R$\&$D was started in 2013, the production was launched in 2016, the first installation grade soldered assembly was produced in 2017, and the production completed in 2021.  Crucial to the R$\&$D process was the availability of a facility which was able to produce small glass-pyrex demonstrators on short timescales for the prototyping phase, as well as a manufacturer for the production which was willing to taken on the challenges and risks of this new technology. As of this date all microchannel assemblies used to produce VELO modules have shown excellent performance, exhibiting no problems due to handling, glueing, wirebonding and thermal cycling.

 Challenges to be considered when deciding whether to implement microchannel technology include considerations of the cost and fragility of the devices, the size of the final coolers and the the effort and skills needed to complete production.  There is extensive ongoing R$\&$D to address the outstanding issues for microchannel production and to make the process more cheap and reliable.  The stringent QA demands on the quality of the bonding can have an impact on the production yield, particularly if very large coolers are produced.  Reducing the size of the cooler should increase the yield, however in this case the connector should be made lower mass and the connection procedure simplified.  This could be addressed by new developments of low mass plug and play hydraulic connections \cite{Francescon:216255} which automatically engage between adjacent modules.  $\rm{R \& D}$ is ongoing and such ‘‘LEGO" connectors have been tested up to 40 bar~\cite{Aglieri:2764386}.  The production of the patterns of channels could be made cheaper, faster, and more flexible with respect to design changes by the use of different techniques.  Maskless laser writing could replace the time consuming and expensive lithographic steps and provide a more prototype-friendly process flow~\cite{mi12091054}.  Alternatively the entire microchannel cooler could be produced with additive manufacturing, based on plastics (acrylic and epoxy-based photosensitive resins) or ceramics (Zirconia, Alumina, Silicon carbide, and Aluminium nitride).  Such a technique was successfully pursued as a back up option for the LHCb VELO, as described in section \cref{sec:alternativecoolingsolutions}.  By using alternative silicon wafer bonding techniques it may be possible to integrate the cooling channels directly into the active silicon and avoid the thermal barrier of the glue layer.  For this, low temperature bonding is being investigated, to enable the etched wafer to be directly bonded to the active silicon, or alternatively the technique of buried microchannels may be used, whereby channels are anisotropically etched into the back of a CMOS sensor in a post processing step.  Finally, for those applications in extremely intense environments where there may be high power consumption in tandem with a need to keep the irradiated sensors at very low temperatures, a move to an alternative coolant such as krypton may be necessary to access lower temperatures for the coolant.  For a more complete discussion of future perspectives in microchannel cooling see~\cite{Mapelli:2712079}.

The LHCb VELO microchannel cooled modules are currently being installed and are expected to run for the next decade.  For the forthcoming Upgrade II \cite{u2phys,u2eoi} new cooling solutions will be needed, among which microchannels incorporating the latest developments will be a strong contender.